\documentclass[letterpaper,12pt]{iopart}
\usepackage{afterpage}
\usepackage{iopams}
\usepackage[dvips]{graphicx}
\usepackage{dcolumn}
\usepackage[usenames]{color}
\usepackage{euscript}
\usepackage{mathrsfs}
\usepackage{enumerate}
\begin{document}

\title[Dependence of test-mass thermal noises on 
beam shape in GW interferometers]{The dependence of test-mass thermal noises on 
beam shape in gravitational-wave interferometers}

\author{Geoffrey Lovelace}
\address{Theoretical Astrophysics, California Institute of Technology,
Pasadena, CA 91125}
\ead{geoffrey@tapir.caltech.edu}

\begin{abstract}
In second-generation, ground-based interferometric gravitational-wave detectors
such as Advanced LIGO, 
the dominant noise at frequencies $f \sim 40$ Hz to $\sim 200$ Hz is expected 
to be due to thermal fluctuations in the mirrors' substrates and coatings which 
induce random fluctuations in the
shape of the mirror face. The laser-light beam averages over these 
fluctuations; the larger the beam and 
the flatter its light-power distribution, the better the averaging and the 
lower the resulting thermal noise.   
In semi-infinite mirrors, scaling laws for the influence of beam shape on the four dominant types of thermal noise 
(coating Brownian, coating thermoelastic, substrate Brownian, and substrate thermoelastic)
have been suggested by various researchers and derived with varying 
degrees of rigour. 
Because these scaling laws are important tools
for current research on optimizing the beam shape, it is important to firm up our
understanding of them.  
This paper (1) gives a summary of the prior
work and of gaps in the prior analyses, (2) gives a unified and rigorous derivation
of all four scaling laws, and (3) explores, relying on work by J. Agresti,
deviations from the scaling laws due to finite mirror size.  
\end{abstract}


\date{\today}


\section{Introduction and Summary}
\label{sec:introduction}
Second generation interferometric gravitational wave detectors such as Advanced LIGO 
will be approximately ten times more sensitive than 
the current LIGO interferometers, leading to an improvement in event rate such 
that the first few hours of data at design sensitivity will contain more signals than the 
entire year-long science run that is presently under way 
\cite{AdvancedLIGOProposal}.
In advanced LIGO's most sensitive frequency band 
(\(f\sim40\mbox{ to }200\mbox{ Hz}\)), the sensitivity is limited by internal 
thermal noise, i.e., by noise in the substrates and reflective coatings of the 
four test masses (see, e.g., figure\ 1 of \cite{Agresti:2004}). Lowering 
the internal thermal noise would increase advanced LIGO's event rate 
throughout that band.

Internal thermal noise can be divided into two different types: 
\emph{Brownian thermal noise} (due to imperfections in the substrate or 
coating material, which couple normal modes of vibration to each other) 
and \emph{thermoelastic noise} (due to random flow of heat in the substrate 
or coating, which causes random thermal expansion). When the laser beam shape 
is Gaussian, the Brownian and thermoelastic noises in the substrate 
(e.g. \cite{liuThorne:2000}) and in the coating 
(e.g. \cite{harry:2002}\cite{BV:2003}) are well understood. 

One way of lowering 
the internal thermal noise is to i) flatten the shape of the laser beam that 
measures the test mass position so it better averages over the mirror 
faces' fluctuating shapes, and ii) enlarge it to the largest size permitted by diffraction losses. 
A specific enlarged, flattened shape, the {\it mesa beam},
has been proposed by O'Shaughnessy and Thorne and explored (theoretically) in
detail by them, d'Ambrosio, Strigin and Vyatchanin \cite{AOT,AOSTV,OSV:2003}, and
by Agresti and DeSalvo \cite{Agresti:2005a,Agresti:2005b}.  The mesa shape was
found to reduce the thermal noise powers by factors of order two, with corresponding
significant increases in the distances to which the planned interferometers can search.
Motivated by this, mesa beams are currently being explored 
experimentally \cite{Agrestietal:2006}\cite{Taralloetal:2007}.

The mesa shape is unlikely to be optimal. Bondarescu and Chen
(Caltech/AEI) are currently seeking the optimal beam shape for each of the four types of noise;
they
are also seeking a balance between the competing demands of the four optimal shapes. Further
research will require balancing practical aspects of mirror design against the (possibly
impractical) ideal shapes.

In all this current research, a crucial tool is 
a set of scaling laws for the dependence of the four types of thermal noise on the
beam shape, in the limit of a mirror that is large compared to the
beam diameter (``semi-infinite mirror'').  These scaling laws have been proposed by
various researchers over the past several years, and they have been derived with varying degrees
of rigour,
and in some cases with unnecessarily restrictive assumptions.  This prior work will be
discussed and critiqued in section\  \ref{sec:prior}.

Because these scaling laws are so important for current research, this
paper scrutinizes them and their accuracies in some detail.
In section \ref{sec:laws} the scaling laws and assumptions underlying them are presented and prior
research on them is described. 
Then in section \ref{sec:Derivation} a unified and rigorous derivation of all four scaling laws is
presented. In section \ref{sec:apps}  the breakdown of the scaling laws due to finite mirror size is explored.  
And finally, in section~\ref{sec:conclusion} a few conclusions are given.

\section{The scaling laws and prior research on them}
\label{sec:laws}

\subsection{Model and Summary}
To explore the effect of the beam shape on the internal thermal noise, 
I consider a cylindrical test mass substrate of radius \(R\) and thickness 
\(H\) and suppose that these size scales are comparable: \(R\sim H\). I choose 
a cylindrical coordinate system \((r,\varphi,z)\) such that \(r=0\) is the 
mirror axis, \(z=0\) is the reflectively coated surface of the mirror 
substrate, and points with \(0<z<H\) are inside the mirror substrate. 

An axisymmetric laser beam with intensity profile \(p(r)\) is normally 
incident on the mirror\footnote{The shape of the mirror faces must also be 
changed slightly (by height changes \(\lesssim\) one wavelength of the laser 
light) so that \(p(r)\) is an eigenmode of the arm cavity. In this paper, I 
assume that the mirror faces take whatever shape is necessary to support a 
beam with intensity profile \(p(r)\).}. The intensity profile is normalized, 
so
\begin{equation}\label{eq:norm}
2\pi \int_0^R {\rmd}r r p(r) = 1.
\end{equation}

The beam measures \(q(t)\), a weighted average of the mirror's longitudinal 
position \(Z(r,\varphi,t)\) (equation (3) of \cite{levin:1998})
\begin{equation}
q(t) \equiv \int_0^{2\pi} \rmd\varphi \int_0^R {\rmd}r r p(r) Z(r,\varphi,t).
\end{equation}
In LIGO, so as to keep diffraction losses \(\lesssim 1\rm{ ppm}\), 
the beam radius over which, say, 95\% of the signal \(q(t)\) is collected, 
is kept significantly smaller than the mirror radius \(R\) and thickness \(H\). 
This motivates the idealization of the mirror as a semi-infinite slab bounded 
by a plane, \(R\rightarrow\infty\), \(H\rightarrow\infty\). (The accuracy of 
this infinite-test-mass (ITM) approximation will be discussed in 
section~\ref{sec:finite}.)

Internal thermal noise will cause small fluctuations in the longitudinal 
position of the mirror \(Z(r,\varphi,t)\). The spectral density 
\(S_q\)
associated with the measurement of the mirror position \(q\) is given by the 
fluctuation dissipation theorem (equation (1) of \cite{levin:1998}):
\begin{equation}\label{eq:FDT}
S_q = \frac{2 k_B T}{\pi^2 f^2} \frac{W_{\rm{diss}}}{F^2}.
\end{equation} Here \(k_B\) is Boltzmann's constant, \(T\) is the temperature 
of the material, and \(W_{\rm{diss}}\) is the power that would be dissipated 
if a longitudinal force \(F\) with frequency \(f\) and pressure distribution 
\(p(r)\) were applied to the mirror surface (Levin's \cite{levin:1998} thought 
experiment). Because the frequencies of interest (i.e. \(f\sim 100\mbox{ Hz}\)) 
are far below the lowest resonant frequencies of the mirror \(f_{\rm{res}} 
\sim \mbox{(a few km/s)} / \mbox{(about 10 cm)} \sim 10^4 \mbox{ Hz}\), the 
hypothetical applied force \(F\) can be idealized as static when computing the 
resulting strain of the mirror. 

Thus the noise \(S_q\) can be computed using the following algorithm: 
\begin{enumerate}
\item Statically deform the (semi-infinite) mirror with a force \(F\) with 
pressure distribution \(p(r)\) the same as the light's intensity profile;  
\item compute the Brownian and thermoelastic dissipated power 
\(W_{\rm{diss}}\) due to the deformation caused by \(F\); 
\item substitute \(W_{\rm{diss}}\) into equation (\ref{eq:FDT}) to get the 
spectral density \(S_q\) of the thermal noise of a measurement of the average 
position \(q\).
\end{enumerate} 

Note that from \(S_q\), one can easily compute the thermally-induced 
gravitational-wave-strain noise power \(S_h(f)\) 
in a measurement by the interferometer. 
If mirrors 1 and 2 are in one arm (of length 
\(L=4 \rm{ km}\)), and mirrors 3 and 4 are in the other arm (also of length 
\(L\)), 
the interferometer
measures \(h\equiv[(q_1-q_2)-(q_3-q_4)]/L\), where 
\(q_{1,2,3,4}\) are the measured positions of the four mirrors.  Because the 
noises in the four test masses are uncorrelated, the spectral density \(S_h\) 
is just \(S_h = (4/L^2) S_q\). In the remainder of this article, when 
referring 
to the noise of a single test mass, the subscript ``q'' will be suppressed 
(i.e. \(S\equiv S_q\)), while the gravitational-wave-strain noise power will 
always be referred to as \(S_h\).

  
In section~\ref{sec:strain}, I compute the strain distribution that results from 
applying the force \(F\) to a homogeneous, isotropic, semi-infinite mirror with 
a very thin reflective coating of a possibly different material. The 
calculation is a straightforward generalization of section~2 of 
\cite{harry:2002}. In this calculation, I model the coating as a thin layer 
(of order microns, as compared to the cm size scales of the substrate) which 
adheres to the mirror surface. In section~\ref{sec:Noise} I use the strain 
distributions to compute each of the four types of thermal noise \(S(f)\). I find that if \(p_1(r)\) and \(p_2(r)\) are two 
different beam shapes, then
\begin{equation}\label{eq:scale}
\frac{S_{1,n}}{S_{2,n}} = \frac{\int_0^\infty {\rmd}k k^n \left[\tilde{p}_1(k)\right]^2}{\int_0^
\infty {\rmd}k k^n 
\left[\tilde{p}_2(k)\right]^2}
\end{equation} where \(n=1\) for coating Brownian and coating thermoelastic 
noise, \(n=0\) for substrate Brownian noise, and \(n=2\) for substrate 
thermoelastic noise. Here
\(\tilde{p}(k)\) is (up to factors of 
\(2\pi\)) the two-dimensional Fourier transform of \(p(r)\) over the surface 
of the mirror:
\begin{eqnarray}
\label{eq:invHT}\tilde{p}(k) & = & \int_0^\infty {\rmd}r r J_0(k r) p(r),\nonumber\\
 p(r) & = & \int_0^\infty {\rmd}k k J_0(k r) \tilde{p}(k).
\end{eqnarray}
Here \(J_0(x)\) is the \(0^{\rm{th}}\) Bessel function of the first kind 
(the axisymmetry allows the 2D Fourier transform to reduce to a 1D Hankel 
transform).

If one knows \(S_{1,n}\), computing \(S_{2,n}\) amounts 
to 
computing simple integrals of \(\tilde{p}(k)\). If one holds everything else 
fixed but changes the beam shape, the scaling law (\ref{eq:scale}) makes it 
straightforward 
to determine the improvement in the thermal noises 
and the corresponding improvement in the interferometer sensitivity.

In the remainder of this paper, I derive these scaling laws, comment on their 
implications for advanced LIGO, and estimate their accuracy for finite test 
masses. In section~\ref{sec:prior}, I discuss prior work related to the
scaling laws. In section~\ref{sec:strain}, I compute the strain \(S_{ij}\) due to 
a hypothetical applied force \(F\) with pressure distribution \(p(r)\). Then, in section~\ref{sec:Noise}, 
I compute the dissipated power \(W_{\rm{diss}}\) for the 
Brownian and thermoelastic dissipation in the coating and the substrate and 
insert \(W_{\rm{diss}}\) into equation (\ref{eq:FDT}) to determine how the noise 
depends on the beam shape. In section~\ref{sec:discussion}, I discuss implications 
of this result for advanced LIGO, and in section~\ref{sec:finite} I discuss the 
accuracies of the infinite-test-mass (ITM) scaling laws by comparing with 
others' finite-test-mass (FTM) predictions for the cases of Gaussian and 
mesa beam shapes. I make some concluding remarks in section~\ref
{sec:conclusion}.

\subsection{Discussion of prior research}\label{sec:prior}

\subsubsection{Thermoelastic substrate noise}

In Levin's thought experiment, the dissipation associated with thermoelastic 
noise arises from heat flow down temperature gradients, which are induced by 
compression of the coating or substrate by the force \(F\). The increase in 
entropy corresponds to a dissipated power.  

In 2003, in connection with
his invention of the mesa beam and exploration of its properties, O'Shaughnessy
used Levin's thought experiment to derive the following scaling law
for the thermoelastic 
substrate noise:
\begin{equation}\label{eq:scaleTEsub}
S_{\rm{sub}}^{\rm{TE}} = C_{\rm{sub}}^{\rm{TE}} \int_0^\infty {\rmd}k k^2 
\tilde{p}^2(k),
\end{equation} where \(C_{\rm{sub}}^{\rm{TE}}\) does not depend on the beam shape.
This scaling law ultimately motivated the other three.  
O'Shaughnessy included his derivation of this law as Appendix H of his (as yet unpublished)
2004 paper with Strigin
and Vyatchanin \cite{OSV:2003} on mesa beams.
He used a slightly different (but no less
rigorous) method from
the unified derivation I give in section \ref{sec:TEsub} [equation (\ref{eq:TESubRes})]. 
O'Shaughnessy wrote the scaling law in terms of
2D Fourier transforms; the reduction to 1D Hankel transforms
makes numerical evaluation of the scaling law (\ref{eq:scaleTEsub}) very efficient (section \ref{sec:discussion}). 

\subsubsection{Thermoelastic coating noise}\label{sec:TECN}

Braginsky and Vyatchanin (Appendix B.2 of \cite{BV:2003}) and Fejer and collaborators 
(section~IV D of 
\cite{fejeretal:2004}) have independently calculated the thermoelastic 
coating noise for Gaussian beam shapes (though the analysis in \cite{BV:2003} is only valid when the coating and
substrate elastic properties are identical \cite{fejeretal:2004}). Scrutinizing the derivation in \cite{fejeretal:2004},
Thorne speculated in 2004 (unpublished) that the thermoelastic coating noise obeys a scaling law of the form 
\begin{equation}
S_{\rm{coat}}^{\rm{TE}} = C_{\rm{coat}}^{\rm{TE}} \int_0^\infty {\rmd}k k 
\tilde{p}^2(k).
\label{eq:TECoat1}
\end{equation}
In 2006 I verified Thorne's conjecture via almost trivial generalizations of the Braginsky-Vyatchanin
and Fejer et.\ al.\ analyses; my derivation is given in section \ref{sec:TEcoat} [equation (\ref{eq:TECoatRes})].  
In 2006 O'Shaughnessy, learning
of my work but not knowing my result, extended a clever dimensional analysis argument 
(section \ref{sec:dimanalysis}) that
he originally invented for Brownian coating noise (below) to the other three types of noises 
\cite{OShaughnessy:2006v1};
but for thermoelastic coating noise he got an answer that disagrees with Thorne's
conjecture and my result (\ref{eq:TECoat1}).  When I pointed out the discrepancy, O'Shaughnessy
found an error in his dimensional analysis and revised his manuscript to give the
correct scaling law (\ref{eq:TECoat1}) \cite{OShaughnessy:2006}.
 
\subsubsection{Brownian coating noise}
In Levin's thought experiment, the dissipation associated with Brownian 
thermal noise can be modelled as arising from a loss angle, which is an 
imaginary (i.e. damping) correction to the material's Young's modulus 
caused by coating or substrate imperfections. 

In 2004, Thorne communicated to O'Shaughnessy and Vyatchanin
his conjecture (\ref{eq:TECoat1}) for the scaling law for thermoelastic coating noise,
and challenged them to find an analogous scaling law for Brownian coating noise.
Independently, they each devised simple arguments that led to the law 
\begin{equation}\label{eq:4}
S_{\rm{coat}}^{\rm{BR}} = C_{\rm{coat}}^{\rm{BR}} \int_0^\infty {\rmd}r r 
p^2(r) = C_{\rm{coat}}^{\rm{BR}} \int_0^\infty {\rmd}k k \tilde{p}^2(k).
\end{equation} O'Shaughnessy gave both an argument
based on dimensional analysis (section \ref{sec:dimanalysis}) and a derivation for the special case that the substrate and coating have the same elastic properties. Vyatchanin's analysis
\cite{Vyatchanin:2004}
was based on a derivation for Gaussian beams, followed by an argument that,
if a beam with another shape $p(r)$ can be constructed by superposing Gaussian
beams, then this scaling law must hold also for that other shape.  

The scaling law (\ref{eq:4}) is {\it local}, i.e., the noise at a point on the mirror depends only on the beam intensity {\it evaluated at that point}. Thorne's intuition, however, led him to believe (incorrectly) that the scaling law should be {\it nonlocal}\footnote{It turns out (section \ref{sec:BRcoat}) that nonlocal terms {\it do} appear at intermediate steps in the derivation but {\it do not} contribute to the scaling law itself.}. 
Consequently, Thorne was so highly
sceptical of O'Shaughnessy's and Vyatchanin's arguments and the claimed scaling law that he --- unfortunately --- dissuaded
both O'Shaughnessy and Vyatchanin from publishing their arguments and result. 

The following year (2005), Thorne, still sceptical of the O'Shaughnessy-Vyatchanin
result  (\ref{eq:4}), suggested to me that I carry out a detailed derivation of
the Brownian-coating-noise scaling law from first principles.  My analysis, based on
Levin's method and reported in
this paper, gave the result (\ref{eq:4}), in agreement with O'Shaughnessy and
Vyatchanin, and motivated O'Shaughnessy to publish \cite{OShaughnessy:2006} his dimensional-analysis
argument. 

O'Shaughnessy's derivation is restricted 
(unrealistically) to identical elastic properties for substrate and coating.  
My derivation [equation (\ref{eq:BRCoatRes}) below] allows the substrate and
the coating to have different elastic properties.   Vyatchanin's derivation is valid
only for those beam shapes that can be achieved by superposing Gaussians --- though
it might well be that any shape can be achieved in this way.
My derivation is definitely valid for any axially symmetric beam shape \(p(r)\).

\subsubsection{Brownian substrate noise}

In 2005 Vinet proposed \cite{vinet:2005} the following scaling law for the substrate
Brownian noise:
\begin{equation}\label{eq:scaleBRsub}
S_{\rm{sub}}^{\rm{BR}} = C_{\rm{sub}}^{\rm{BR}} \int_0^\infty {\rmd}k 
\tilde{p}^2(k).
\end{equation} 
He deduced this law as a trivial consequence of his equations (1) -- (3).  He did not
present a derivation of those equations, but he recognized 
that they can be obtained by generalizing the derivation in \cite{BHV}, which
assumes that the beam shape is Gaussian. 
In section \ref{sec:BRsub}, I explicitly derive equation (\ref{eq:scaleBRsub}).  In parallel
with my work, O'Shaughnessy applied his dimensional analysis technique to verify
Vinet's scaling law (\ref{eq:scaleBRsub}).


\subsubsection{Dimensional analysis}\label{sec:dimanalysis}
O'Shaughnessy's dimensional analysis argument, referred to above, consists of three steps: 
\begin{enumerate}
\item The scaling laws must take
the form of a translation-invariant inner product of \(p(r)\) with itself, since the
mirror is taken to be semi-infinite. In the Fourier domain, 
for axisymmetric beam shapes \(p_1(r)\) and \(p_2(r)\), the scaling law must then take the form:
\begin{equation}
\frac{S_1}{S_2} =\frac{\int_0^\infty \rmd k k \tilde{G}(k) \tilde{p}_1^2(k)}{\int_0^\infty \rmd k k \tilde{G}(k) \tilde{p}_2^2(k)}.
\end{equation} 
\item The only length scale (other than the width of the beam) is the small
coating thickness \(d\), so  \(\tilde{G}(k)= k^n d\) for coating thermal noise and 
\(\tilde{G}(k)=k^n\) for substrate thermal noise.
\item The power \(n\) is chosen by demanding that, when the beam shape is a Gaussian,
the noise scale as the correct power of the beam width.
\end{enumerate}

This argument turns out to produce the correct scaling laws, but without sufficient care, it can also
lead one amiss.
For instance, when considering thermoelastic coating noise, step (ii) must be amended, 
since there \emph{is} a second
length scale: the characteristic length of diffusive heat flow \cite{BV:2003}\cite{fejeretal:2004}. 
In his original manuscript \cite{OShaughnessy:2006v1}, O'Shaughnessy neglected this second length scale, and incorrectly
deduced that \(n=3\) for coating thermoelastic noise.  After I contacted O'Shaughnessy regarding this error, he corrected his analysis  \cite{OShaughnessy:2006} and obtained the same result, \(n=1\), as
I had derived (section  \ref{sec:TEcoat}) below.

\section{Derivation of the infinite-test-mass (ITM) scaling laws}
\label{sec:Derivation}
\subsection{Strain of a semi-infinite body with thin facial coatings due to a 
static, axisymmetric force}\label{sec:strain}
The thermal noise is determined by the symmetric part of the strain \(S_{ij}\) 
that the test mass would experience if a normal force with pressure \(p(r)\) 
were applied to the mirror surface. In this section, I evaluate \(S_{ij}\) in 
the mirror substrate and coating. In section~\ref{sec:Noise}, I use these results 
to compute \(W_{\rm{diss}}\) [which, by equation (\ref{eq:FDT}), determines the 
thermal noise].

If the displacement vector of an element of the test mass is \(u_i\), then the 
strain \(S_{ij}\) is \(S_{ij} = \nabla_j u_i\). Following the methods developed 
in \cite{BHV} (but correcting some typographical errors), equation (19) of
\cite{liuThorne:2000} gives the cylindrical components of the 
displacement of the test mass substrate:
\numparts
\begin{eqnarray}\label{eq:ur}
u_r & = & \frac{1}{2\mu} \int_0^\infty {\rmd}k J_1(k r) \rme^{-k z} \left(1-\frac
{\lambda+2\mu}{\lambda+\mu}+kz\right)\tilde{p}(k),\nonumber\\ \\
u_\varphi & = & 0,\\\label{eq:uz}
u_z & = & \frac{1}{2\mu} \int_0^\infty {\rmd}k J_0(k r) \rme^{-k z} \left(1+\frac{\mu}
{\lambda+\mu}+kz\right) \tilde{p}(k).\nonumber\\
\end{eqnarray}
\endnumparts Here \(\lambda\) and \(\mu\) are the Lam\'e coefficients 
of the substrate. The vector 
\(u_i\) satisfies the equilibrium equation \(\nabla_j T_{ij} = 0\). 
(Throughout this paper, I use the Einstein summation convention.)

The non-vanishing components of the symmetric part of the strain are 
[with commas denoting partial derivatives]
\numparts
\begin{eqnarray}
\theta & = & S_{ii},\label{eq:expansion}\\
S_{rr} & = & u_{r,r} = \theta - S_{zz} - S_{\varphi\varphi},\\
S_{\varphi\varphi} & = & \frac{u_r}{r},\\
S_{(rz)} & = & S_{(zr)} = \frac{1}{2}(u_{r,z}+u_{z,r}),\\
S_{zz} & = & u_{z,z}\label{eq:uz,z}.
\end{eqnarray}
\endnumparts Evaluating the derivatives of 
equations (\ref{eq:ur}) -- (\ref{eq:uz}) and inserting the result into 
equations (\ref{eq:expansion}) -- (\ref{eq:uz,z}) gives
\numparts
\begin{eqnarray}\label{eq:thetaSub}
\theta & = & \frac{1}{2\mu}\int_0^\infty {\rmd}k k J_0(k r) 
\left(\frac{-2\mu}{\lambda+\mu}\right) \rme^{-k z} \tilde{p}(k),\nonumber\\ \\
S_{rr} & = & \theta - S_{zz} - S_{\varphi\varphi},\nonumber\\ \\
\label{eq:SphiphiSub}S_{\varphi\varphi} & = & \frac{1}{2\mu} 
\int_0^\infty {\rmd}k \frac{J_1(k r)}{r} \rme^{-k z} 
\left(1-\frac{\lambda+2\mu}{\lambda+\mu}+k z\right)\tilde{p}(k),\nonumber\\ \\
S_{(zr)} & = & -\frac{1}{2 \mu} \int_0^\infty {\rmd}k k J_1(k r) (k z) \rme^{-k z} 
\tilde{p}(k),\nonumber\\ \\
\label{eq:SzzSub} S_{zz} & = & \frac{1}{2\mu} \int_0^\infty {\rmd}k k J_0(k r) 
\left(-\frac{\mu}{\lambda+\mu}-kz\right) \rme^{-kz} \tilde{p}(k).\nonumber\\
\end{eqnarray}
\endnumparts Setting \(z=0\) in 
equations (\ref{eq:thetaSub})--(\ref{eq:SzzSub}) and combining with 
equation (\ref{eq:invHT}) yields the nonvanishing stresses on the substrate 
surface:
\numparts
\begin{eqnarray}\label{eq:allExp}
\left. \theta\right|_{z=0} & = & \left(\frac{-1}{\lambda+\mu} \right) p(r), \\
\label{eq:rexpand} S_{rr}\left.\right|_{z=0} &=& 
\frac{1}{2}\left(\frac{-1}{\lambda+\mu}\right) p(r) 
- S_{\varphi\varphi}\left.\right|_{z=0},\\
S_{\varphi\varphi}\left.\right|_{z=0} & = & \frac{1}{2} 
\left(\frac{-1}{\lambda+\mu}\right) \int_0^\infty {\rmd}k 
\frac{J_1(k r)}{r}\tilde{p}(k),\\
\left. S_{zz} \right|_{z=0} & = & 
\frac{1}{2}\left(\frac{-1}{\lambda+\mu}\right)p(r).\label{eq:allExp2}
\end{eqnarray}
\endnumparts Here I have used the identity 
\begin{equation}
\int_0^\infty {\rmd}k k J_0(k r) J_0(k r^\prime) = 
\frac{\delta(r^\prime - r)}{r^\prime}.
\end{equation}
Note that on the surface of the substrate \(\theta\) and \(S_{zz}\) are 
\emph{local} [i.e. their values at any point depend only on the value of 
\(p(r)\) at that point], while \(S_{\varphi\varphi}\) is \emph{nonlocal}. 
The component \(S_{rr}\) can be written as the sum of a local part and a 
nonlocal part; the nonlocal part of \(S_{rr}\) is 
just \(-S_{\varphi\varphi}\).

The thin coating approximation gives the nonvanishing components of the 
coating 
strain in terms of the strain on the substrate surface (equation (A4) of 
\cite{harry:2002}):
\numparts
\begin{eqnarray}\label{eq:thetaCoat}
\theta^{\rm{coat}} & = &  
\frac{\lambda+2\mu_{\rm{coat}}}{\lambda_{\rm{coat}}
+2\mu_{\rm{coat}}}\left.\left(\theta - S_{zz} \right)\right|_{z=0} 
\nonumber\\
& & + \frac{\lambda+2\mu}{\lambda_{\rm{coat}}+2 \mu_{\rm{coat}}} 
\left.S_{zz}\right|_{z=0},\\
\label{eq:srrelim}S_{rr}^{\rm{coat}} & = & 
\left. S_{rr}\right|_{z=0} = \theta^{\rm{coat}} 
- S_{\varphi\varphi}^{\rm{coat}}-S_{zz}^{\rm{coat}},\nonumber\\ \\
S_{\varphi\varphi}^{\rm{coat}} & = & 
\left. S_{\varphi\varphi}\right|_{z=0},\nonumber\\ \\
S_{zz}^{\rm{coat}} & = & 
\frac{\lambda-\lambda_{\rm{coat}}}{\lambda_{\rm{coat}}
+2\mu_{\rm{coat}}}\left.\left(\theta - S_{zz} \right)\right|_{z=0} 
\nonumber\\
& & + \frac{\lambda+2\mu}{\lambda_{\rm{coat}}+2 \mu_{\rm{coat}}} 
\left.S_{zz}\right|_{z=0}.\label{eq:SzzCoat} \nonumber \\
\end{eqnarray}
\endnumparts In \cite{harry:2002}, these conditions are said to 
hold in the limit that the Poisson ratio of the substrate and coating are 
``not too different,'' but this restriction is unnecessary 
(see \ref{app:junction}).

Finally, after inserting equations (\ref{eq:allExp}) -- (\ref{eq:allExp2}) into 
equations (\ref{eq:thetaCoat}) -- (\ref{eq:SzzCoat}) 
I conclude that \(\theta^{\rm{coat}}\) and \(S_{zz}^{\rm{coat}}\) are 
\emph{local}; while \(S_{\varphi\varphi}^{\rm{coat}}\) and 
\(S_{rr}^{\rm{coat}}\) are \emph{nonlocal}. However, this nonlocality turns 
out \emph{not} to influence the coating noises. This is because, after using 
equation (\ref{eq:srrelim}) to eliminate \(S_{rr}^{\rm{coat}}\), it turns out that 
the remaining nonlocal part \(S_{\varphi\varphi}^{\rm{coat}}\) only appears in 
the coating \(W_{\rm{diss}}\) [according to equations (\ref{eq:WdissBRCoat}) 
and (\ref{eq:TECoat})] via the integral 
\begin{eqnarray}
\fl \int_0^\infty {\rmd}r r S_{(ij)}^{\rm{coat}} S_{(ij)}^{\rm{coat}} 
 =  \int_0^\infty {\rmd}r r \Big[(S_{rr}^{\rm{coat}})^2 
+ (S_{\varphi\varphi}^{\rm{coat}})^2 + (S_{zz}^{\rm{coat}})^2\Big] \nonumber\\
\fl =  \int_0^\infty {\rmd}r r 
\Big[\left(\theta^{\rm{coat}}-S_{zz}^{\rm{coat}}\right)^2 
+  (S_{zz}^{\rm{coat}})^2
+  2 (S_{\varphi\varphi}^{\rm{coat}})^2 - 2 S_{\varphi\varphi}^{\rm{coat}} 
\left(\theta^{\rm{coat}}-S_{zz}^{\rm{coat}}\right)\Big].
\end{eqnarray}
In \ref{sec:nonlocalCancels}, I show that
\begin{eqnarray}\label{eq:apRes}
\int_0^\infty {\rmd}r r (S_{\varphi\varphi}^{\rm{coat}})^2 
- S_{\varphi\varphi}^{\rm{coat}} \left(\theta^{\rm{coat}}
-S_{zz}^{\rm{coat}}\right) = 0,
\end{eqnarray} so only the local parts of the strain 
(\(\theta^{\rm{coat}}\) and \(S_{zz}^{\rm{coat}}\)) influence the 
thermal noise.
This fact turns out to imply local coating scaling laws in agreement with 
O'Shaughnessy's \cite{OShaughnessy:2006} and Vyatchanin's 
\cite{Vyatchanin:2004} arguments (section~\ref{sec:Noise}).

\subsection{Internal thermal noise}
\label{sec:Noise}
\subsubsection{Brownian coating noise}\label{sec:BRcoat}

For Brownian thermal noise in an elastic material, the dissipated power 
is [equation (12) of \cite{levin:1998} with a static applied force and 
with \(U=-(1/2)S_{ij}T_{ij}\)]
\begin{eqnarray}\label{eq:WdissBRCoatTemp}
W_{\rm{diss}} = - \pi f \int_0^d {\rmd}z \int_0^{2\pi} \rmd\varphi \int_0^\infty {\rmd}r r 
\phi(f) S_{ij} T_{ij}.
\end{eqnarray} Here \(\phi\) is the loss angle (i.e., the imaginary, 
damping part of the Young's modulus of the coating material) and \(T_{ij}\) 
is the stress. When the material is the thin reflective coating of a mirror, 
there are effectively \emph{two} loss angles \cite{harry:2002}, \(\phi_{\|}\) 
and \(\phi_{\bot}\), defined so that in the previous equation
\begin{eqnarray}\label{eq:twoloss}
\phi(f) S_{ij} T_{ij} & \rightarrow & 
\phi_{\|}(f)\left(S_{rr}^{\rm{coat}} T_{rr}^{\rm{coat}} 
+ S_{\varphi\varphi}^{\rm{coat}} T_{\varphi\varphi}^{\rm{coat}} \right) 
+ \phi_{\bot}(f) S_{zz}^{\rm{coat}} T_{zz}^{\rm{coat}}\nonumber\\
&  = & \phi_{\|}(f) S_{ij}^{\rm{coat}} T_{ij}^{\rm{coat}} 
+ \left(\phi_{\bot}-\phi_{\|}\right) S_{zz}^{\rm{coat}} T_{zz}^{\rm{coat}}.
\end{eqnarray} This result can be obtained by combining 
equations (4) and (13) -- (15) of \cite{harry:2002} with equation (9) of 
\cite{levin:1998} and recalling that in the coating, the strain 
[equations (\ref{eq:thetaCoat}) -- (\ref{eq:SzzCoat})] is 
diagonal.

For a homogeneous coating, the stress \(T_{ij}^{\rm{coat}}\) is
\begin{equation}\label{eq:Tij}
T_{ij}^{\rm{coat}} = -\lambda^{\rm{coat}} \theta^{\rm{coat}} \delta_{ij} 
- 2 \mu^{\rm{coat}} S_{(ij)}^{\rm{coat}},
\end{equation} where \(\lambda^{\rm{coat}}\) and \(\mu^{\rm{coat}}\) 
are the Lam\'{e} coefficients of the coating, \(S_{(ij)}^{\rm{coat}}\) 
is the symmetric part of the coating strain, and \(\theta \equiv S_{ii}\) 
is the expansion. Combining equations (\ref{eq:twoloss}), 
(\ref{eq:WdissBRCoatTemp}) and (\ref{eq:Tij}) gives the following expression 
for \(W_{\rm{diss}}\): 
\begin{eqnarray}\label{eq:WdissBRCoat}
W_{\rm{diss}} & = & 2 \pi^2 f d \phi_{\|}(f) \int_0^\infty {\rmd}r r A  
 + 2 \pi^2 f d \left[\phi_{\bot}(f) - \phi_{\|}(f)\right] 
\int_0^\infty {\rmd}r r  B,\nonumber\\
A & = & \left(\lambda^{\rm{coat}} \theta^2_{\rm{coat}}
+ 2 \mu^{\rm{coat}} S_{(ij)}^{\rm{coat}} S_{(ij)}^{\rm{coat}} \right),
\nonumber\\
B & = & S_{zz}^{\rm{coat}}\left(\lambda^{\rm{coat}} \theta^{\rm{coat}} 
+ 2 \mu^{\rm{coat}} S_{zz}^{\rm{coat}} \right).
\end{eqnarray}

Combining equations (\ref{eq:WdissBRCoat}), (\ref{eq:thetaCoat}) -- 
(\ref{eq:SzzCoat}), (\ref{eq:allExp}) -- (\ref{eq:allExp2}), and 
(\ref{eq:apRes}) and then inserting the result into equation (\ref{eq:FDT}) 
gives the spectral density \(S\) of the Brownian coating noise. However, for 
the present purpose, only terms involving the beam shape are relevant. 
Absorbing all other terms into a single constant \(C_{\rm{coat}}^{\rm{BR}}\) 
yields
\begin{equation}\label{eq:BRCoatRes}
S_{\rm{coat}}^{\rm{BR}} 
= C_{\rm{coat}}^{\rm{BR}} \int_0^\infty {\rmd}r r p^2(r).
\end{equation} This is a local scaling law; i.e., the noise at each point on 
the mirror's surface is proportional to the square of the beam intensity there. 
This law is the same as O'Shaughnessy's \cite{OShaughnessy:2006} and 
Vyatchanin's \cite{Vyatchanin:2004} scaling law for the Brownian coating 
thermal noise.

Parseval's equation [which follows from equation (\ref{eq:invHT})] makes it easy 
to rewrite this scaling law in the Fourier domain, which will facilitate 
comparison with the substrate noise. The result is
\begin{equation}\label{eq:coatBRRes}
S_{\rm{coat}}^{\rm{BR}} = C_{\rm{coat}}^{\rm{BR}} \int_0^\infty {\rmd}k k 
\tilde{p}^2(k).
\end{equation}
\subsubsection{Thermoelastic Coating Noise}
\label{sec:TEcoat}
The calculation of the thermoelastic coating noise is similar to the 
calculation of Brownian coating noise. But now, in response to the static, 
normal applied pressure \(p(r)\), the dissipated power \(W_{\rm{diss}}\) is 
caused by heat flow, \(\propto \nabla \delta T\), down a temperature gradient 
\(\nabla \delta T\) caused by the material's deformation:
\begin{equation}\label{eq:temppert}
W_{\rm{diss}} = \frac{\pi\kappa}{T} \int_0^\infty {\rmd}z \int_0^\infty {\rmd}r r 
\left(\nabla \delta T\right)^2.
\end{equation} [equation (5) of \cite{liuThorne:2000} in the case of a 
static applied force and after evaluating the time average and trivial 
\(\varphi\) integral]. Here \(T\) is the temperature of the coating in the 
absence of the deformation and \(\kappa\) is the material's coefficient of 
thermal conductivity.

Braginsky and Vyatchanin \cite{BV:2003} and Fejer and collaborators 
\cite{fejeretal:2004} have independently solved for the thermoelastic coating 
noise. The results obtained in \cite{BV:2003} are correct only when the 
coating and substrate have the same elastic properties 
(section~I in \cite{fejeretal:2004}); however, this restriction is not 
relevant here, since \cite{fejeretal:2004} and \cite{BV:2003} agree on 
the coating thermoelastic noise's dependence on the beam shape \(p(r)\).

If the temperature change were adiabatic, \(\delta T\) would simply be 
proportional to \(\theta^{\rm{coat}}\) (see, e.g., equation (12) of 
\cite{liuThorne:2000}). (Physically, this simply means that the 
temperature of an element in the coating changes linearly with volume.) 
However, as noted in \cite{BV:2003}, the diffusive heat characteristic 
length \(\ell_D\) of the substrate and coating (on the order of mm) is far 
larger than the coating thickness \(d\) (which is on the order of a few 
microns). Because diffusive heat flow in the longitudinal direction 
is \emph{not} negligible, heat flow in the direction normal to the coating 
cannot be treated adiabatically \cite{BV:2003}. By contrast, the 
\emph{substrate} thermoelastic noise \emph{can} be treated adiabatically 
(section~\ref{sec:TEsub}), as can the heat flow in the plane of the coating 
(``tangential'' heat flow).

Because the tangential heat flow is adiabatic, 
\(\partial\delta T/\partial r  \sim \theta/w\), where \(w\sim\rm{ cm}\) is 
the length scale over which \(p(r)\) varies. On the other hand, 
\(\partial \delta T/\partial z \sim \theta/\ell_D\), where 
\(\ell_D\sim\rm{ mm}\) is the diffusive heat characteristic length. 
Because the tangential derivatives are much smaller than the longitudinal 
derivatives, all derivatives except \(\partial/\partial z\) may be neglected. 
It follows that \(W_{\rm{diss}}\) will depend only on \(p(r)\) and not on its 
radial derivatives.

Based on these observations, Braginsky and Vyatchanin \cite{BV:2003} and Fejer 
and collaborators \cite{fejeretal:2004} solve the thermoconductivity 
equation (e.g., equation~(1) of \cite{fejeretal:2004}) for the temperature 
perturbations \(\delta T\). Both \cite{BV:2003} and \cite{fejeretal:2004}
assume that the beam shape is Gaussian, but it is quite easy
to generalize their arguments to non-Gaussian beam shapes.  
Combining equations (B5) -- (B7), (66), and (68) of 
\cite{fejeretal:2004} (but now regarding their function \(\rho(r)\) as a generic beam shape)
shows that the temperature perturbations in the coating have the form
\begin{equation}\label{eq:tempprop}
\delta T \propto p(r) \times F(z),
\end{equation} where \(F(z)\) is a function of \(z\) only. [Equivalently, equation~(\ref{eq:tempprop})
can be obtained by combining equations (B.10) and (B.12) 
of \cite{BV:2003} (but now regarding \(\theta\) as an expansion corresponding to a generic beam shape) with equations (\ref{eq:allExp}) and (\ref{eq:thetaCoat}).] The precise form of 
\(F(z)\) is given in \cite{BV:2003} and \cite{fejeretal:2004} but is not 
needed in the present discussion.

Next, Braginsky and Vyatchanin compute the squared gradient 
\((\nabla \delta T)^2 \simeq (\partial \delta T/\partial z)^2\) in 
equation (\ref{eq:temppert}) to obtain \(W_{\rm{diss}}\); Fejer and collaborators 
instead compute \(W_{\rm{diss}}\) by considering the interaction of i) the 
\emph{unperturbed} stress and strain [i.e., the stress and strain due to 
\(p(r)\) when temperature perturbations are neglected], and ii) the (complex) 
\emph{perturbations} of the stress and strain caused by the small temperature 
perturbations \(\delta T\). Both methods lead to the following expression for 
\(W_{\rm{diss}}\): (equations (B.13) and (B.10) of \cite{BV:2003}; equation (69) 
of \cite{fejeretal:2004})
\begin{equation}\label{eq:TECoat}
W_{\rm{diss}} = {\rm{const}} \times \int_0^\infty {\rmd}r r p^2(r).
\end{equation} Plugging this result into equation (\ref{eq:FDT}) gives the scaling 
law
\begin{equation}\label{eq:TECoatRes}
S_{\rm{coat}}^{\rm{TE}} = C_{\rm{coat}}^{\rm{TE}} \int_0^\infty {\rmd}r 
r p^2(r) = C_{\rm{coat}}^{\rm{TE}} \int_0^\infty {\rmd}k k \tilde{p}^2(k).
\end{equation} This is the same scaling law as for Brownian coating thermal 
noise. The coating thermoelastic noise is local and is the same as 
O'Shaughnessy's \cite{OShaughnessy:2006} and Vyatchanin's 
\cite{Vyatchanin:2004} law for Brownian coating thermal noise.

\subsubsection{Brownian Substrate Noise}\label{sec:BRsub}
For Brownian substrate thermal noise there is only one relevant loss angle, 
\(\phi\), so the dissipated power is (equation (49) of \cite{liuThorne:2000} 
with a static applied force)
\begin{eqnarray}\label{eq:WdissBRSubTemp}
W_{\rm{diss}} = 2 \pi^2 f \phi(f) \int_0^\infty {\rmd}z \int_0^\infty {\rmd}r r  
\left(\lambda \theta^2 + 2 \mu S_{(ij)} S_{(ij)}\right).\nonumber\\
\end{eqnarray} The integral of the squared strain can be expanded as
\begin{eqnarray}\label{eq:WdissSubExpand}
\int_0^\infty {\rmd}z \int_0^\infty {\rmd}r r S_{(ij)} S_{(ij)} =   
\int_0^\infty {\rmd}z \int_0^\infty {\rmd}r r \left( S_{rr}^2 + S_{\varphi\varphi}^2 
+ S_{zz}^2\right) \nonumber\\
 =  \int_0^\infty {\rmd}z \int_0^\infty {\rmd}r r \big[\left(\theta-S_{zz}\right)^2 
+  S_{zz}^2 + 2 S_{(rz)}^2 + 2 S_{\varphi\varphi}^2 
- 2 S_{\varphi\varphi} \left(\theta-S_{zz}\right)\big]. 
\end{eqnarray} 
In \ref{sec:nonlocalCancels}, I show that
\begin{eqnarray}\label{eq:apResSub}
\int_0^\infty {\rmd}r r \left[S_{\varphi\varphi}^2 
- S_{\varphi\varphi}\left(\theta-S_{zz}\right)\right] = 0.
\end{eqnarray} Substituting this result into equation (\ref{eq:WdissSubExpand}) 
yields
\begin{eqnarray}\label{eq:WdissBRSub}
\fl W_{\rm{diss}}  =   2 \pi^2 f \phi(f) \int_0^\infty {\rmd}z \int_0^\infty {\rmd}r r 
\Big[\lambda \theta^2 + 2\mu\left(\theta-S_{zz}\right)^2 
+    2 \mu S_{zz}^2 + 4 \mu S_{(rz)}^2\Big].
\end{eqnarray} This expression can be evaluated term by term. Inserting 
equation (\ref{eq:thetaSub}) into the integral of \(\theta^2\) gives
\begin{eqnarray}\label{eq:wide2}
\fl I_\theta \equiv \int_0^\infty {\rmd}z \int_0^\infty {\rmd}r r  \theta^2 \nonumber\\ \fl = 
\frac{1}{4 \mu^2} \left(\frac{2\mu}{\lambda+\mu}\right)^2\int_0^\infty {\rmd}k k 
\tilde{p}(k) \int_0^\infty {\rmd}k^\prime k^\prime \tilde{p}(k^\prime) \int_0^\infty 
{\rmd}z \rme^{-(k+k^\prime) z} \int_0^\infty {\rmd}r r  J_0(k r) J_0(k^\prime r). 
\end{eqnarray}
Using the identity 
\begin{equation}
\int_0^\infty {\rmd}r r J_n(k r) J_n(k^\prime r) = \frac{\delta(k-k^\prime)}{k}
\end{equation} on equation (\ref{eq:wide2}) and evaluating the integral over \(z\) 
yields
\begin{eqnarray}
I_\theta & =  & \frac{1}{8 \mu^2} 
\left(\frac{2\mu}{\lambda+\mu}\right)^2\int_0^\infty {\rmd}k \tilde{p}^2(k).
\end{eqnarray} The other terms in equation (\ref{eq:WdissBRSub}) can be evaluated 
similarly; they all turn out to have the same dependence on \(\tilde{p}(k)\) 
as \(I_\theta\). Inserting this result for \(W_{\rm{diss}}\) into 
equation (\ref{eq:FDT}) gives the scaling law
\begin{equation}\label{eq:BRSubRes}
S_{\rm{sub}}^{\rm{BR}} = C_{\rm{sub}}^{\rm{BR}} \int_0^\infty {\rmd}k 
\tilde{p}^2(k).
\end{equation} This scaling law is the same as the scaling law (\ref{eq:coatBRRes}) for the coating 
thermal noise except that the \(z\) integration has 
reduced the power of \(k\) in the integrand by one. This scaling law agrees 
with equations (1) -- (2) of \cite{vinet:2005}.

\subsubsection{Thermoelastic Substrate Noise}
\label{sec:TEsub}
In contrast to the case of coating thermoelastic noise, the substrate 
thermoelastic noise \emph{can} be treated using the adiabatic approximation. 
Therefore, the temperature perturbations \(\delta T\) that drive the substrate 
thermoelastic noise \(S_{\rm{sub}}^{\rm{TE}}\) are proportional to the 
expansion, i.e. \(\delta T \propto \theta\). This implies (e.g., by 
equation (\ref{eq:temppert}), or equation (13) of \cite{liuThorne:2000}) 
\begin{equation}\label{eq:TESub}
S_{\rm{sub}}^{\rm{TE}} = C_{\rm{sub}}^{\rm{TE}} \int_0^\infty {\rmd}z 
\int_0^\infty {\rmd}r r (\nabla \theta)^2
\end{equation} with \(C_{\rm{sub}}^{\rm{TE}}\) independent of the 
strain (and thus also the beam shape). Inserting equation (\ref{eq:thetaSub}) into 
equation (\ref{eq:TESub}) gives the scaling law; after absorbing all constants 
into \(C_{\rm{sub}}^{\rm{TE}}\), it takes the form 
\begin{equation}\label{eq:TESubRes}
S_{\rm{sub}}^{\rm{TE}} = C_{\rm{sub}}^{\rm{TE}} \int_0^\infty {\rmd}k k^2 
\tilde{p}^2(k),
\end{equation} which O'Shaughnessy, Strigin, and Vyatchanin obtain in 
\cite{OSV:2003}. This scaling law is the same as the scaling law (\ref{eq:BRSubRes}) for the 
substrate Brownian noise except that the gradient 
raises the power of \(k\) by two.

\section{Applying the ITM scaling laws to second-generation gravitational-wave interferometers}\label{sec:apps}
To illustrate the scaling laws (\ref{eq:BRCoatRes}), (\ref{eq:TECoatRes}), 
(\ref{eq:BRSubRes}), and (\ref{eq:TESubRes}), suppose that the noise 
\(S_{\tau,k}\) [with beam shape \(p_k(r)\) and thermal noise type \(\tau\)] is 
known. Here and throughout the remainder of this article, \(\tau\) is a label 
that takes one of the following values: coating Brownian (Coat BR), coating 
thermoelastic (Coat TE), substrate Brownian (Sub BR), or substrate 
thermoelastic (Sub TE).

Now, if the beam shape were changed to \(p_u(r)\) [while holding 
everything\footnote{Since here I am neglecting edge effects, ``everything'' 
means the temperature, the materials' elastic and thermal properties, the 
coating thickness, and the frequency. In section~\ref{sec:finite}, when edge 
effects are considered, it will be the diffraction loss, not the mirror size, 
that is held fixed.} else fixed], then the unknown noise \(S_{\tau,u}\) [with 
beam shape \(p_u(r)\)] would be [equation (\ref{eq:scale})]:
\begin{eqnarray}\label{eq:scalegen}
S_{\tau,u} = C_{\rm{ITM}}^2[\tau;p_u,p_k] S_{\tau,k}, 
\end{eqnarray} with
\begin{equation}\label{eq:cdef}
C_{\rm{ITM}}^2[\tau;p_u,p_k] \equiv \frac{\int_0^\infty {\rmd}k k^{n(\tau)} 
\tilde{p}^2_u(k)}{\int_0^\infty {\rmd}k k^{n(\tau)} \tilde{p}_k^2(k)}
\end{equation} and
\begin{eqnarray}\label{eq:ndef}
n(\tau) \equiv \left\{ \begin{array}{r@{\quad:\quad}l} 
1 & \tau=\rm{Coat BR or Coat TE} \\ 0 & \tau=\rm{Sub BR} \\
2 & \tau=\rm{Sub TE}
\end{array} \right.
\end{eqnarray} When the beam shape is changed from \(p_k\) to \(p_u\), the 
amplitude 
sensitivity changes by a factor of \(C_{\rm{ITM}}[\tau;p_u,p_k]\).

\subsection{Implications for advanced LIGO}
\begin{figure}[t]
\centerline{\includegraphics[width=8cm]{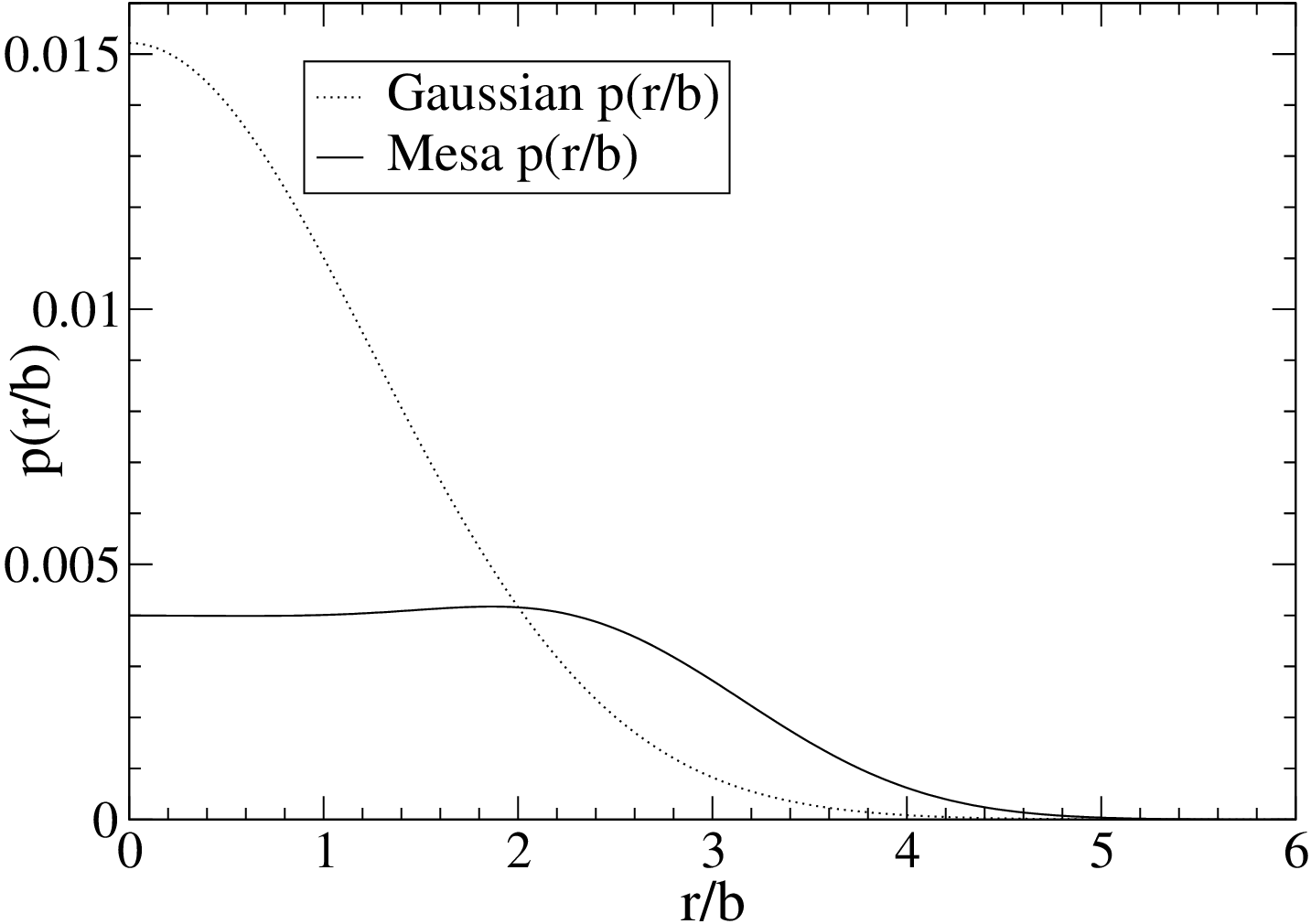}}
\caption{A plot of \(p_{\rm{gauss}}(r/b)\) and \(p_{\rm{mesa}}(r/b)\) for beams 
with 1 ppm diffraction losses (in the clipping approximation) on a mirror with 
radius \(R=17\rm{ cm}\). Here \(b=\sqrt{L\lambda/2\pi}=2.6\rm{ cm}\) is the 
width of the smallest Gaussian beam that can resonate in a LIGO arm cavity with 
length \(L=4\rm{ km}\) and light wavelength \(\lambda=1064\rm{ nm}\). 
\label{fig:shape}}
\end{figure}
\label{sec:discussion}
In advanced LIGO, the thermal noise may be significantly reduced by 
changing the 
shape of the laser beam. One proposal is to replace the Gaussian beam 
shape with a mesa 
beam (also called a flat-top beam) \cite{AOT}. O'Shaughnessy, 
Strigin, and Vyatchanin 
\cite{OSV:2003} have calculated the resulting reduction in 
substrate thermoelastic 
noise, 
Vinet has done the same for substrate Brownian thermal 
noise \cite{vinet:2005} and 
Agresti \cite{Agresti:2004} and Agresti and DeSalvo 
\cite{Agresti:2005a,Agresti:2005b} 
have done the same for both substrate and coating thermal noises---
all for the realistic 
case of finite mirrors. The reduction in thermal noise can also be 
understood as a 
consequence of the simple ITM scaling laws derived above. Although 
I only compare 
Gaussian and flat-top beams here, the scaling law given in 
equation (\ref{eq:scale}) 
makes 
it simple---if one neglects finite-test-mass (FTM) effects---
to compute the relative 
change in sensitivity for any two beam shapes.

The normalized Gaussian beam shape is
\begin{equation}\label{eq:pgauss}
p_{\rm{gauss}}(w;r) = \frac{\rme^{-r^2/w^2}}{\pi w^2}
\end{equation} where \(w\) is the width of the Gaussian beam. It is 
straightforward to compute \(\tilde{p}_{\rm{gauss}}(w;k)\), since the integral 
can be done analytically; the result is
\begin{equation}
\tilde{p}_{\rm{gauss}}(w;k) = \int_0^\infty {\rmd}r r J_0(k r) 
\frac{\rme^{-r^2/w^2}}{\pi w^2} = \frac{1}{2\pi} \rme^{-k^2 w^2/4}.
\end{equation}

\begin{figure}
\centerline{\includegraphics[width=8cm]{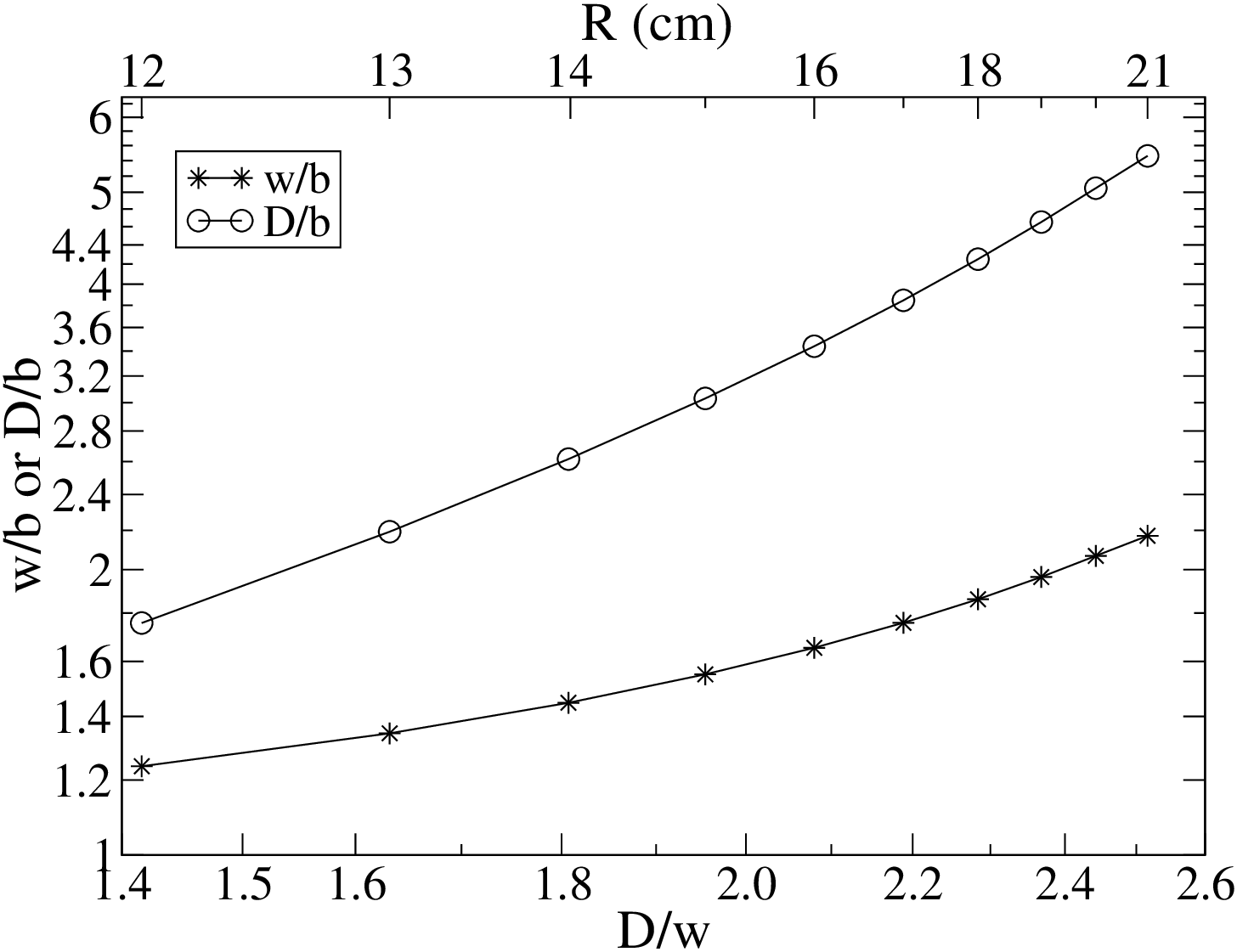}}
\caption{A log-log plot of the Gaussian beam-width parameter \(w\) and mesa 
beam-width parameter \(D\) as a function of mirror radius \(R\) 
(top of figure), for mirrors with 1 ppm diffraction loss in the clipping 
approximation. The ratio \(D/w\) is shown on the bottom of the figure. The 
parameter \(b\) is defined in figure\ \ref{fig:shape}.}
\label{figxcoords}
\end{figure}

In position space, the mesa beam can be written as (equation (2.5) of 
\cite{AOSTV})
\begin{eqnarray}\label{eq:pmesa}
\fl p_{\rm{mesa}}(D;r) & = & N \Bigg| 2\pi \int_0^D {\rmd}r^\prime r^\prime 
\exp\left[-\frac{(r^2+r^{\prime2})(1-\rmi)}{2 b^2}\right] 
\times I_0\Big[\frac{r r^\prime (1-\rmi)}{b^2}\Big]\Bigg|^2.
\end{eqnarray} Here \(D\) is a measure of the width of the beam, 
\(b \equiv \sqrt{\lambda L/2\pi}\), with \(L=4\rm{ km}\) the arm length 
and \(\lambda=1064\rm{ nm}\) the wavelength of the laser beam's primary 
frequency, and \(N\) is a normalization constant adjusted so 
equation (\ref{eq:norm}) is satisfied. Note that \(p_{\rm{mesa}}(r)\) must be 
evaluated numerically; to compute \(\tilde{p}(k)\) efficiently, I use the Fast Hankel 
Transform algorithm \cite{Siegman:1977}. 

Examples of the Gaussian and mesa shapes are plotted in figure\ \ref{fig:shape}. 
In figure\ \ref{figxcoords}, the width parameters \(w\) and \(D\) of a 
sequence\footnote{This particular sequence was chosen to facilitate comparison 
with the results of \cite{Agresti:2005b}, which includes edge effects.} 
of Gaussian and mesa beams are plotted as a function of mirror radius \(R\) for 
beams with 1 ppm of diffraction loss in the clipping 
approximation\footnote{In the clipping approximation, the diffraction loss is 
simply \(2\pi\int_R^\infty {\rmd}r r p(r)\), where \(R\) is the mirror radius. In 
the ITM approximation, \(R\) is larger than all other length scales; however, 
the actual, finite value of \(R\) must be used in the clipping approximation 
for the diffraction loss to be nonvanishing.}. The ratio \(D/w\) is also shown 
on the bottom horizontal axis. It is sometimes useful to regard \(D\) 
and \(w\) (for 1 ppm losses) as functions of \(D/w\) rather than of \(R\) --- 
with \(D/w\) actually being a surrogate for \(R\).

The following three cases use equations (\ref{eq:scalegen}) -- (\ref{eq:ndef}) to 
illustrate how the thermal noise in advanced LIGO changes with different 
choices 
of Gaussian and mesa beam shapes.
\subsubsection{Noise of a resized Gaussian beam}\label{sec:ResizeGauss}
\begin{figure}
\centerline{\includegraphics[width=8cm]{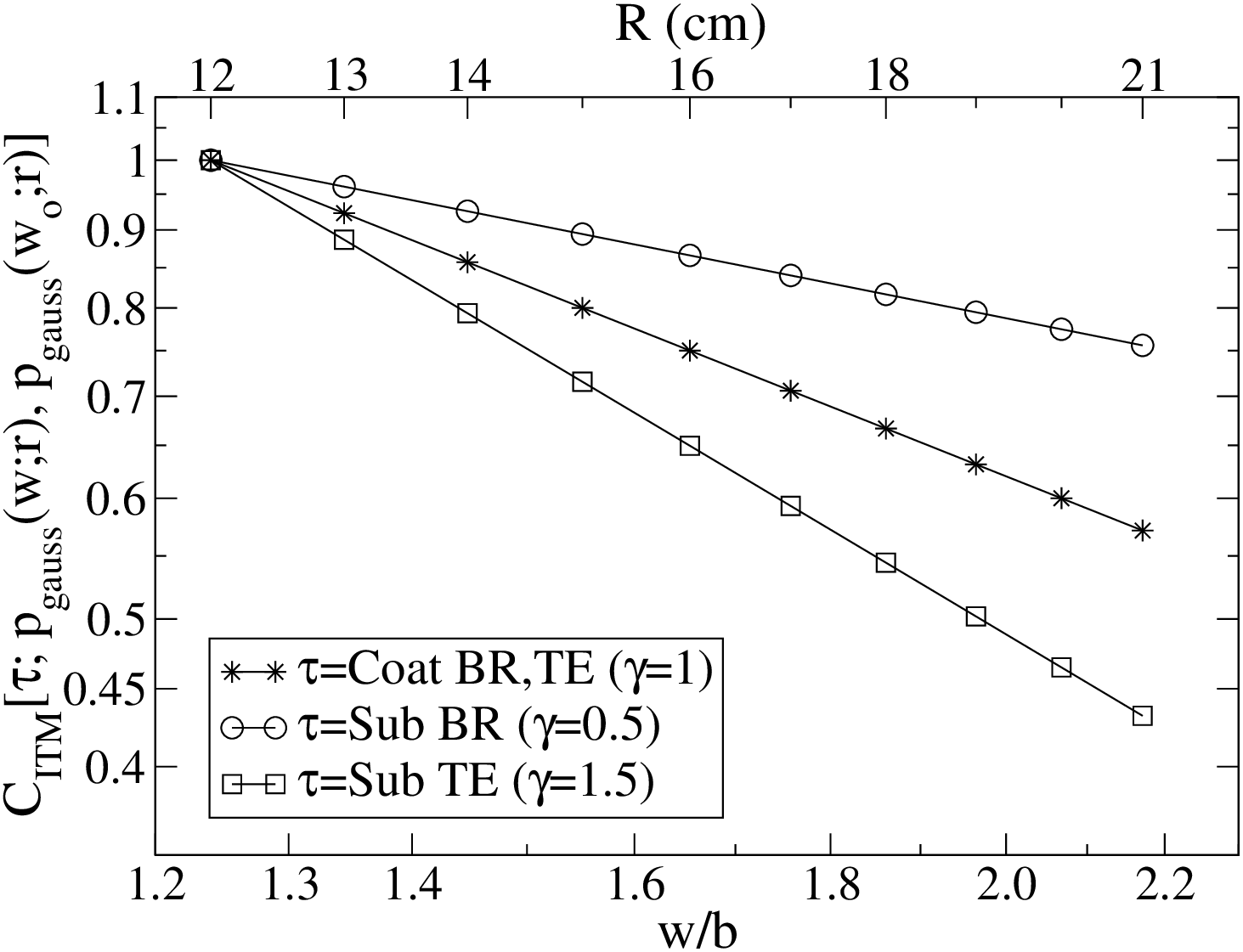}}
\caption{The scaling of thermal noises with beam width \(w\) for Gaussian beams 
in the infinite-test-mass (ITM) approximation. More specifically: a log-log 
plot 
of \(C_{\rm{ITM}}[\tau;p_{\rm{gauss}}(w;r),p_{\rm{gauss}}(w_o;r)]\) as a 
function of \(w/b\). Here \(w_o/b = 1.24\), which corresponds to 
\(R=12\rm{ cm}\) and 1 ppm diffraction losses. Each curve is a power law 
obeying \(C\propto 1/w^\gamma\).}
\label{figResizeGauss}
\end{figure}
Suppose \(p_k(r) = p_{\rm{gauss}}(w_o;r)\). Then the thermal noises for a 
Gaussian beam of some different size \(w\) are determined by evaluating 
\(C_{\rm{ITM}}[\tau;p_{\rm{gauss}}(w;r),p_{\rm{gauss}}(w_o;r)]\) 
[equation (\ref{eq:cdef})] and inserting the result into equation (\ref{eq:scalegen}). 
In this well-known case (see, e.g., the discussion and references in 
\cite{Agresti:2005b}), \(C_{\rm{ITM}}\) can be evaluated analytically, 
yielding the following relation:
\begin{eqnarray}
C_{\rm{ITM}}^2[\tau;p_{\rm{gauss}}(w;r),p_{\rm{gauss}}(w_o;r)] \propto 
\frac{1}{w^{n(\tau)+1}}.
\end{eqnarray}
In figure~\ref{figResizeGauss}, 
\(C_{\rm{ITM}}[\tau;p_{\rm{gauss}}(w;r),p_{\rm{gauss}}(w_o;r)]\) is plotted as 
a function of the beam width \(w\). 
\subsubsection{Noise of a resized mesa beam}\label{sec:ResizeMesa}
\begin{figure}
\centerline{\includegraphics[width=8cm]{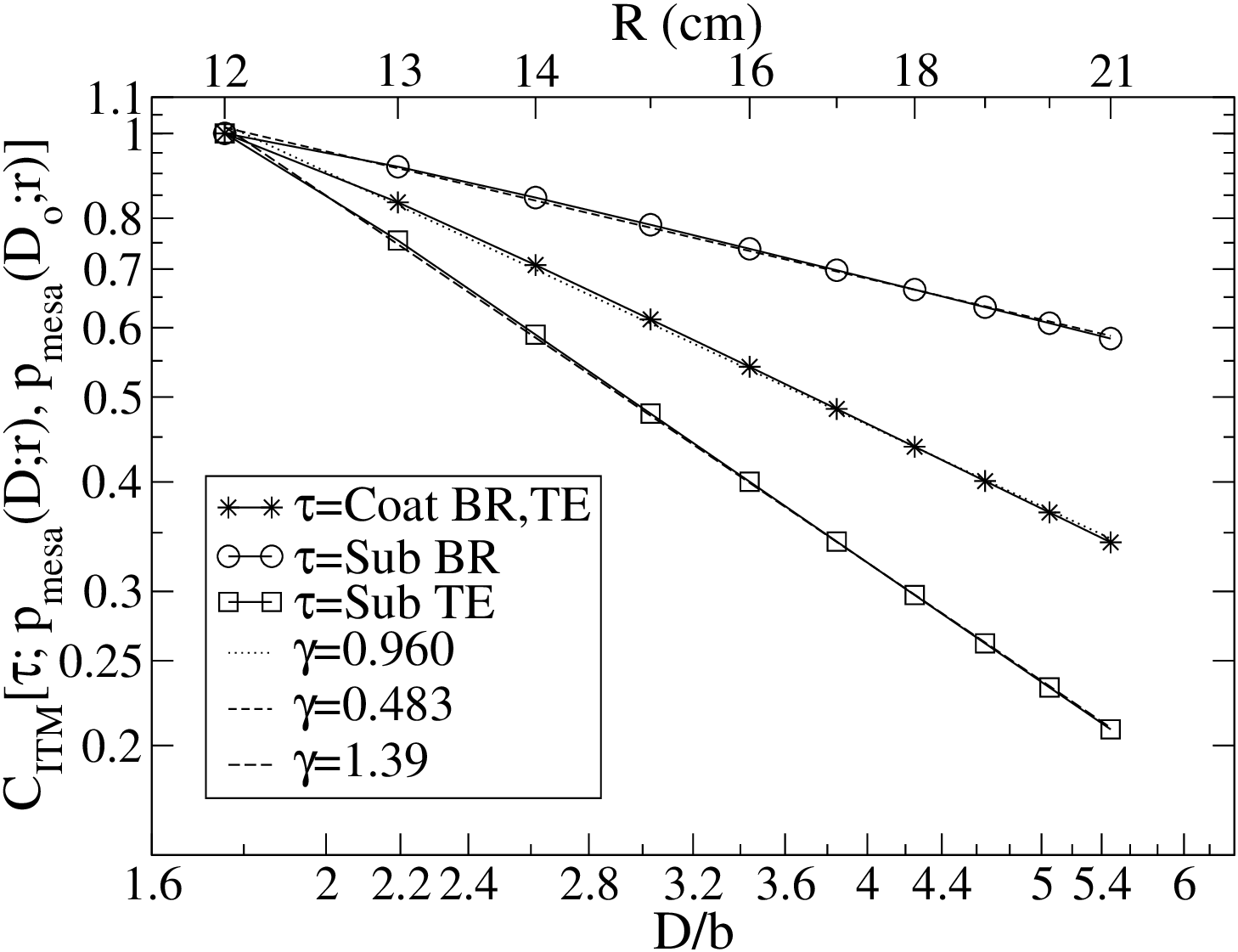}}
\caption{The scaling of thermal noises with beam-width parameter \(D\) for 
mesa beams in the infinite-test-mass approximation. More specifically: 
a log-log plot of 
\(C_{\rm{ITM}}[\tau;p_{\rm{mesa}}(D;r),p_{\rm{mesa}}(D_o;r)]\) as a function 
of \(D/b\). Here \(D_o/b=1.76\), which corresponds to a mirror radius of 
\(12\rm{ cm}\) and 1 ppm diffraction losses. The curves are well approximated 
by power laws of the form \(C\propto 1/D^\gamma\).}
\label{figResizeMesa}
\end{figure}
Suppose \(p_k(r)=p_{\rm{mesa}}(D_o;r)\). Then the thermal noises for a mesa 
beam of some different size \(D\) are determined by evaluating 
\(C_{\rm{ITM}}[\tau;p_{\rm{mesa}}(D;r),p_{\rm{mesa}}(D_o;r)]\) 
[equation (\ref{eq:cdef})] and inserting the results into equation (\ref{eq:scalegen}). 
As shown in figure~\ref{figResizeMesa}, in this case \(C_{\rm{ITM}}\) does not 
scale as an exact power of \(D\) (although the actual relations are very well 
approximated by power laws).

\subsubsection{Noise reduction by switching from a Gaussian beam to a mesa beam 
with the same diffraction loss and mirror radius}\label{sec:GaussToMesa}

Finally, the scaling law (\ref{eq:scalegen}) can be used to estimate the reduction in 
thermal noise by switching from a Gaussian beam to a mesa beam that has the 
same clipping-approximation diffraction loss on a mirror of the same radius. 

Two complications in the resized-beam scalings are not present when scaling 
from Gaussian to mesa beams. First, while the original and resized beams were 
associated with different-sized mirrors, now the Gaussian and mesa beams are 
associated with the \emph{same} mirror. Second, when relating the Gaussian and 
mesa beams, there is no need to specify a fiducial beam size (i.e. there is no 
analogue of \(w_o\) and \(D_o\)). Without these two complications, the 
Gaussian-to-mesa scaling is perhaps conceptually cleaner than the resized-beam 
scalings.

\begin{figure}
\centerline{\includegraphics[width=8cm]{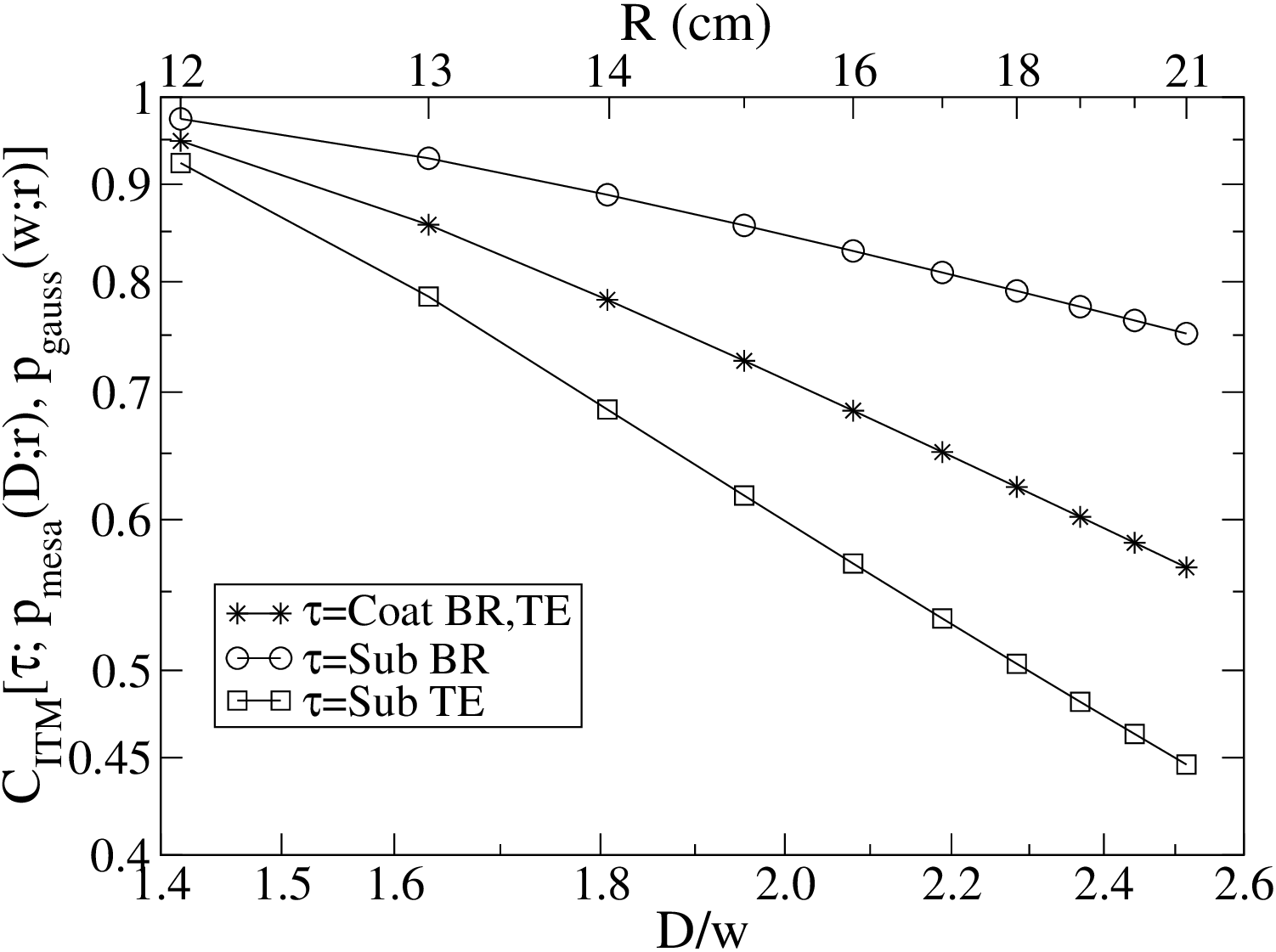}}
\caption{The improvement in amplitude sensitivity when mesa beams are used 
instead of Gaussian beams. More specifically: a log-log plot of 
\(C_{\rm{ITM}}[\tau;p_{\rm{mesa}}(D;r),p_{\rm{gauss}}(w;r)]\) as a function of 
\(D/w\). For each mirror radius \(R\), \(w\) and \(D\) are chosen so that the 
diffraction losses are 1 ppm in the clipping approximation.}
\label{figGaussToMesa}\label{fig:1}
\end{figure}

Figure \ref{figGaussToMesa} shows 
\(C_{\rm{ITM}}[\tau;p_{\rm{mesa}}(D;r),p_{\rm{gauss}}(w;r)]\) for the sequence 
of beams shown in figure\ \ref{figxcoords} (beams with 1 ppm diffraction loss in 
mirrors of the same radius \(R\)). The relative improvement in amplitude 
sensitivity increases monotonically with the mirror radius \(R\), or 
equivalently, with \(D/w\); however, when edge effects 
(i.e. finite-test-mass effects) are included, there is a limit to how much the 
sensitivity can be improved (section~\ref{sec:finite}).

\subsection{Errors due to neglecting finite-test-mass (FTM) effects}
\label{sec:finite}
\begin{figure}
\centerline{\includegraphics[width=8cm]{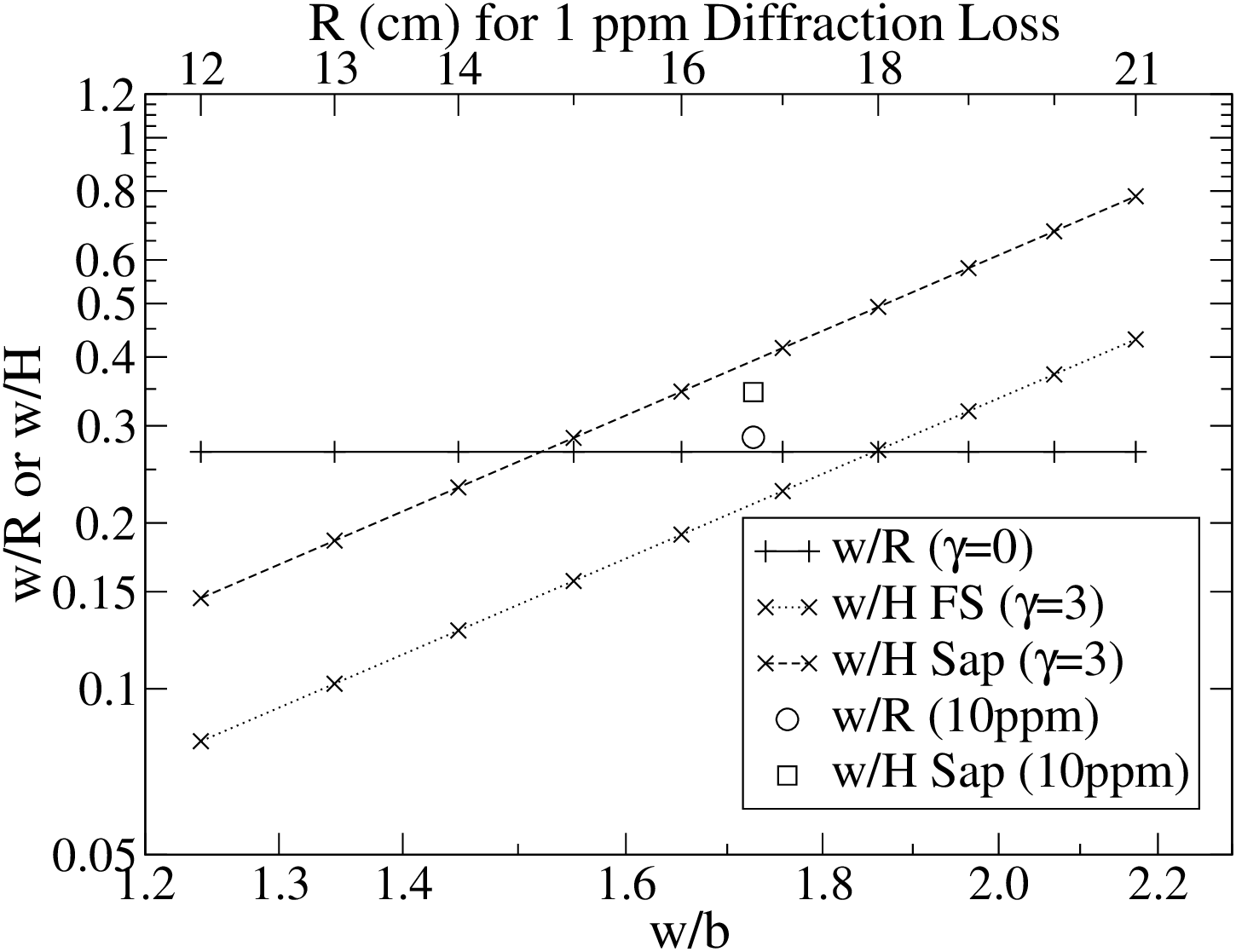}}
\caption{How the Gaussian beam width parameter \(w\) compares to the mirror 
radius \(R\) and thickness \(H\), when i) the radius \(R\) is fixed so the 
clipping-approximation diffraction loss is 1 ppm (unless a 10 ppm loss is 
indicated), and ii) the thickness \(H\) is then determined by holding the mass 
at 40 kg, the advanced-LIGO baseline mirror mass. Each curve is proportional 
to \(w^\gamma\). FS and Sap mean fused-silica and sapphire substrates.}
\label{figDimlessGauss}
\end{figure}
\begin{figure}
\centerline{\includegraphics[width=8cm]{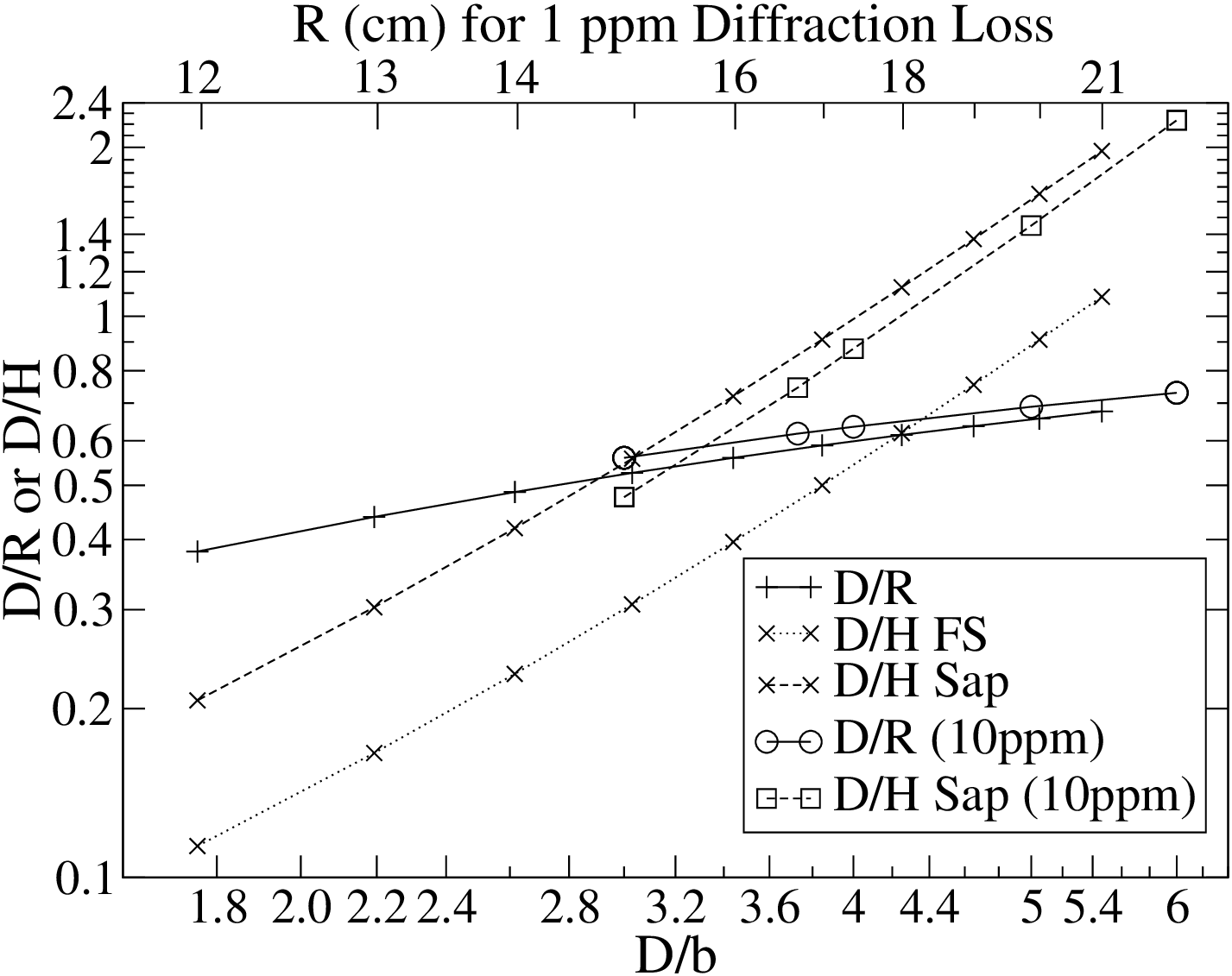}}
\caption{How the mesa beam width parameter \(D\) compares to the mirror 
radius \(R\) and thickness \(H\), when i) the radius \(R\) is fixed so the 
clipping-approximation diffraction loss is 1 ppm (unless a 10 ppm loss is 
indicated), and ii) the thickness \(H\) is then determined by holding the mass 
at 40 kg. The mirror radius \(R\) for 1 ppm losses is shown on the top axis; 
the 10 ppm mirror radii are (from left to right) 
\(R^{\rm{10 ppm}}=\)13.94 cm, 15.7 cm, 16.37 cm, 18.85 cm, and 21.36 cm. 
FS and Sap mean fused-silica and sapphire substrates.}
\label{figDimlessMesa}
\end{figure}
In the previous section, the ITM scaling laws predicted that, if the 
diffraction losses are held fixed, then the coating and substrate noises 
decrease monotonically with increasing beam width [figures \ref{figResizeGauss}, 
\ref{figResizeMesa}, and \ref{figGaussToMesa}]. In other words, for a given 
diffraction loss, the optimal beam width is simply ``as large as possible.'' 

However, this conclusion is only as strong as the ITM approximation. Its 
validity can be checked by comparing the beam widths to the corresponding 
mirror dimensions. In our modelling, the mirror radii \(R\) are adjusted to 
maintain a constant clipping-approximation diffraction loss (CADL) 
[figure\ \ref{figxcoords}], while the thicknesses \(H\) is then determined by 
requiring the mirror mass be 40 kg---the design specification for Advanced 
LIGO. (Thus \(H\) will depend on whether the substrate is Fused Silica (FS) or 
sapphire (Sap), since the densities of these materials differ by a factor of 
about 2.) 

As shown in figures \ref{figDimlessGauss} and \ref{figDimlessMesa}, for the 
sequences of beam widths considered in section~\ref{sec:discussion}, \(w\) and 
\(D\) can approach or even exceed \(H\) while simultaneously being significant 
fractions of the \(R\). Consequently, edge effects (finite test-mass effects) 
may significantly change the sensitivity scalings depicted in 
figures \ref{figResizeGauss}, \ref{figResizeMesa}, and \ref{figGaussToMesa}.

To estimate the importance of these edge effects, I compare the results in 
sections\ \ref{sec:ResizeGauss} -- \ref{sec:GaussToMesa} to the finite-test-mass 
(FTM) 
results\footnote{The FTM data used here assume that the coating extends 
all the way to 
the edge of the substrate face. In advanced-LIGO, the coating radius will 
actually be 
several mm smaller than the substrate radius (the baseline substrate radius for 
advanced 
LIGO is 170\rm{ mm}).} of Agresti and DeSalvo \cite{Agresti:2005b} 
(all types of 
thermal noise, 1 ppm CADL) and O'Shaughnessy, Strigin, and Vyatchanin 
\cite{OSV:2003} 
(substrate thermoelastic noise only, 10 ppm CADL). Specifically, 
from these data I read 
off the ratio
\begin{equation}
C_{\rm{FTM}}[\tau;p_u(r),p_k(r)] \equiv 
\sqrt{\frac{S^{\rm{FTM}}_{\tau,u}}{S^{\rm{FTM}}_{\tau,k}}}.
\end{equation} This change in sensitivity can be compared to 
\(C_{\rm{ITM}}[\tau;p_k(r),p_u(r)]\), the change in sensitivity obtained by the 
ITM approximation. Specifically, if
\begin{equation}
\Delta[\tau;p_u(r),p_k(r)] \equiv 
\frac{C_{\rm{FTM}}[\tau;p_u(r),p_k(r)]}{C_{\rm{ITM}}[\tau;p_u(r),p_k(r)]},
\end{equation} then \(|1-\Delta|\) is the fractional error made by using the 
ITM approximation to compute \(C[\tau;p_u(r),p_k(r)]\).

In the following subsections, I consider the errors \(|1-\Delta|\) made 
[sections \ref{sec:ResizeGauss} -- \ref{sec:GaussToMesa}] by neglecting FTM 
effects.
\subsubsection{Resized Gaussian beam}
\begin{figure}
\centerline{\includegraphics[width=8cm]{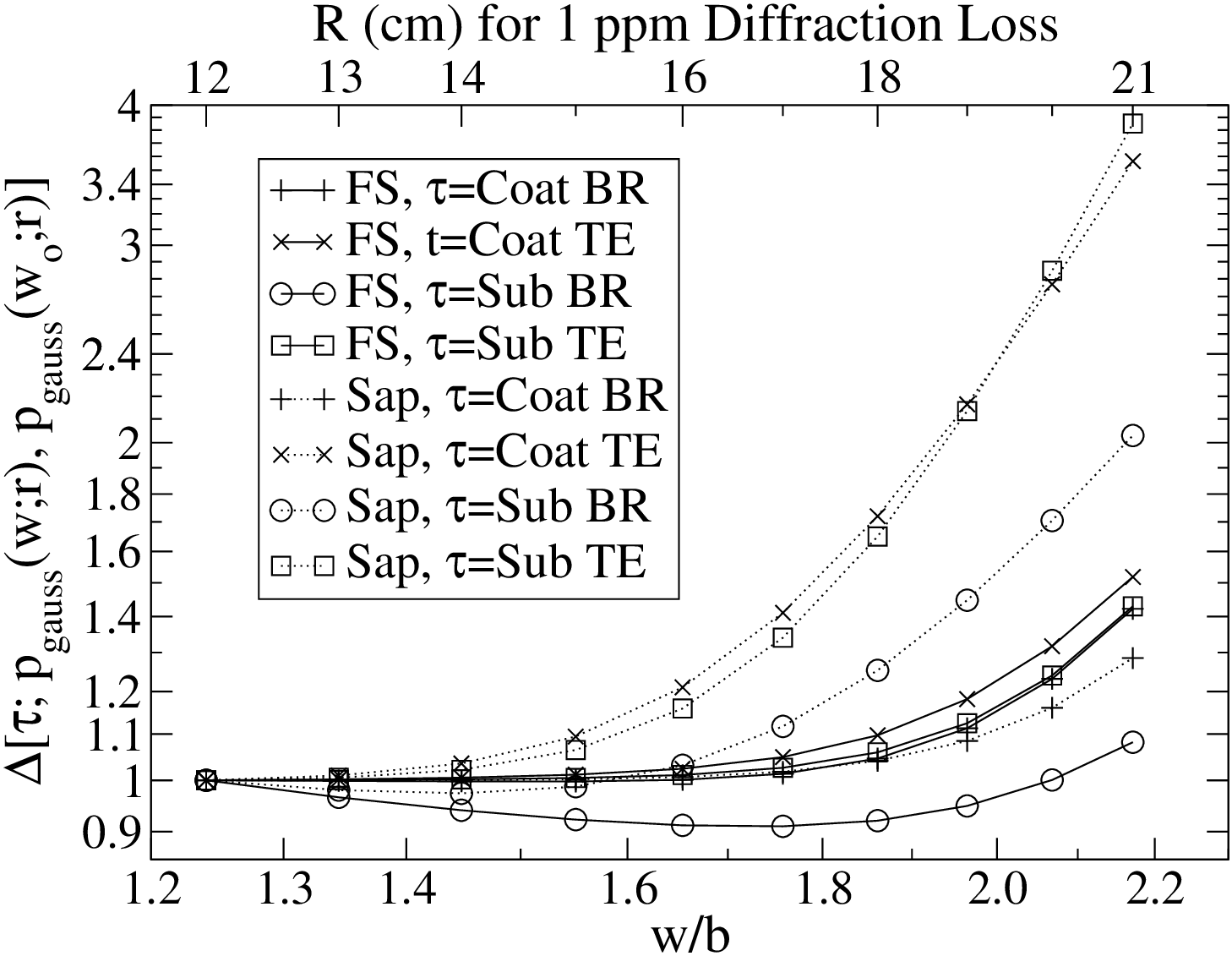}}
\caption{A log-log plot of 
\(\Delta[\tau;p_{\rm{gauss}}(w;r),p_{\rm{gauss}}(w_o;r)]\). 
The fractional error of the sensitivity change made by neglecting edge effects 
is 
\(|1-\Delta|\). Here \(w_o/b = 1.24\), which corresponds to \(R=12\rm{ cm}\) 
and 
1 ppm diffraction losses. The FTM values are obtained by taking ratios of the 
noises 
calculated by Agresti and DeSalvo \cite{Agresti:2005b}. FS and Sap mean 
fused-silica 
and sapphire substrates.}
\label{figCFTMResizeGauss}
\end{figure}
Figure \ref{figCFTMResizeGauss} plots 
\(\Delta[\tau;p_{\rm{gauss}}(w;r),p_{\rm{gauss}}(w_o;r)]\) for mirror 
substrates made of fused silica, the baseline material for advanced LIGO 
mirrors \cite{AdvancedLIGOProposal}. For comparison, the figure also shows the 
corresponding values of \(\Delta\) for sapphire substrates.

When the substrate is fused silica, the ITM and FTM scaling laws agree to 
better than about 10\% so long as \(R\lesssim 17\rm{ cm}\), the advanced-LIGO 
baseline mirror radius \cite{AdvancedLIGOProposal}. As \(R\) increases beyond 
about 17 cm, \(|1-\Delta|\) increases dramatically (to about 50\% when 
\(R=21\rm{ cm}\)), because for such large radii the noise \emph{increases} 
(e.g. \cite{Agresti:2005b, OSV:2003}) with \(R\), while the ITM scaling laws 
predict [figure\ \ref{figResizeGauss}] that the noise \emph{always decreases} 
with increasing \(R\).

When the substrate is sapphire, the FTM effects for the thermoelastic noises 
lead to errors that are comparable to the fused-silica FTM errors.  For a 
mirror radius of\footnote{When sapphire was the baseline test-mass material 
for advanced LIGO (it has since been abandoned in favour of fused silica), the 
baseline mirror radius was 
\(R=15.7\rm{ cm}\) \cite{AdvancedLIGOProposalOld}.} \(R=16\rm{ cm}\), the 
fractional error \(|1-\Delta|\) for sapphire substrates is about 15\% for 
substrate thermoelastic noise and about 20\% for coating thermoelastic noise.
\subsubsection{Resized mesa beam}
\begin{figure}
\centerline{\includegraphics[width=8cm]{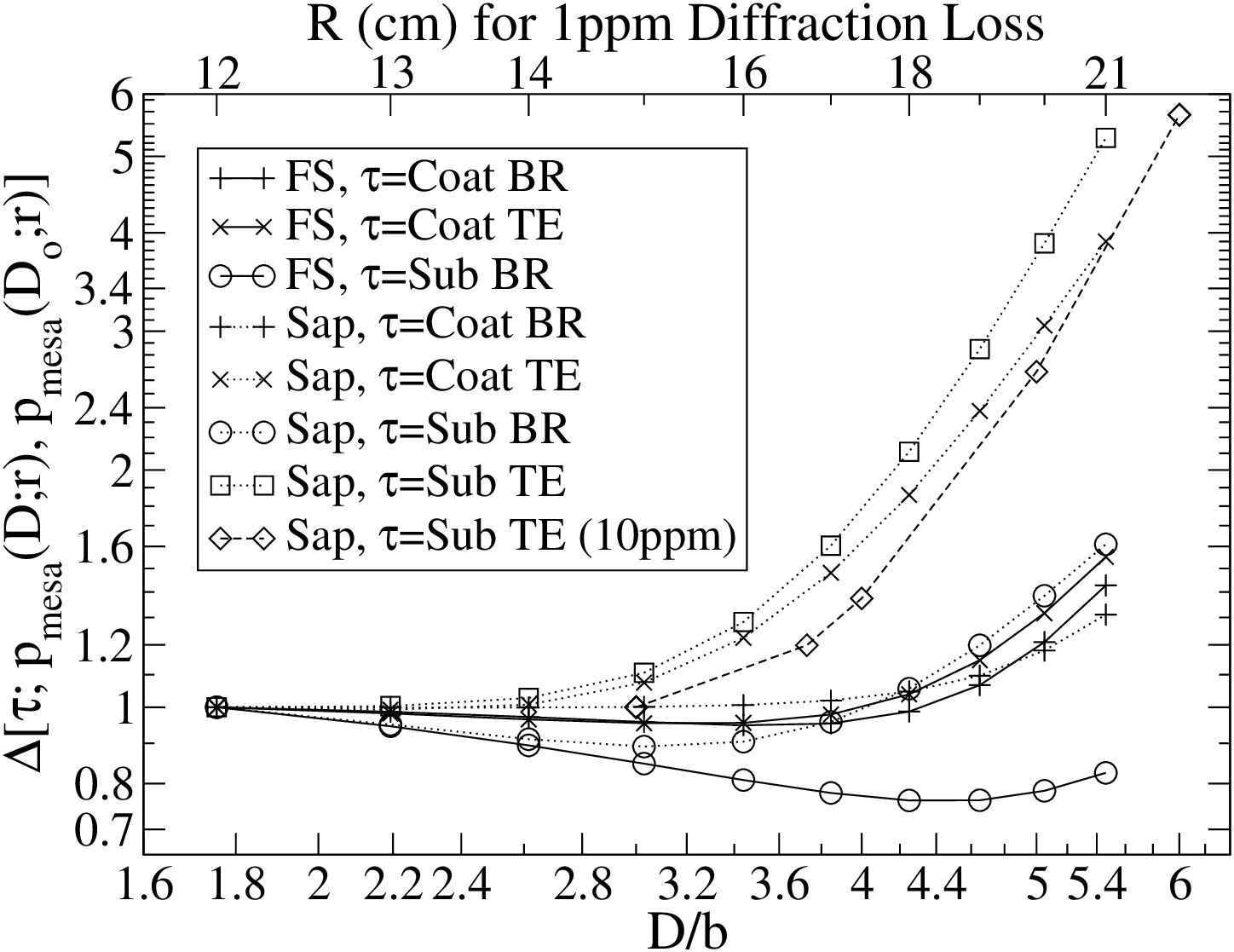}}
\caption{A log-log plot of 
\(\Delta[\tau;p_{\rm{mesa}}(D;r),p_{\rm{mesa}}(D_o;r)]\). The fractional error 
of the sensitivity change made by neglecting edge effects is \(|1-\Delta|\). 
Here the diffraction losses are 1 ppm (unless 10 ppm is indicated), and 
\(D_o/b=1.76\) (\(D_o^{\rm{10 ppm}}=3.00\)), which corresponds to a mirror 
radius \(R=12\rm{ cm}\) (\(R^{\rm{10 ppm}}=13.94\rm{ cm}\)). The 
corresponding mirror radii are given on the top axis (1 ppm losses) and in 
figure\ \ref{figDimlessMesa} (10 ppm losses). The FTM values are obtained by 
taking ratios of the noises calculated by Agresti and DeSalvo 
\cite{Agresti:2005b}, except for the 10 ppm values due to O'Shaughnessy, 
Strigin, and Vyatchanin \cite{OSV:2003}. FS and Sap mean fused-silica and 
sapphire substrates. (The fused-silica substrate thermoelastic noise is 
negligible; this case is omitted from the figure.)}
\label{figCFTMResizeMesa}
\end{figure}

The FTM effects in the resized-mesa-beam case are similar to the 
resized-Gaussian-beam FTM effects. Figure \ref{figCFTMResizeMesa} 
plots \(\Delta[\tau;p_{\rm{mesa}}(D;r),p_{\rm{mesa}}(D_o;r)]\). When the 
substrate is fused silica and \(R\lesssim 17\rm{ cm}\), the ITM scaling law 
errs by less than about 10\% for the coating noises and by less than about 25\% 
for the substrate Brownian noise. (The substrate thermoelastic noise is 
negligible when the substrate is fused silica \cite{Agresti:2005b}.) Again, the 
ITM scaling law disagrees more and more strongly as \(R\) is increased 
beyond \(17\rm{ cm}\). In this regime, the noise \emph{increases} with \(R\), 
but the ITM scaling law [figure\ \ref{figResizeMesa}] predicts that the noise 
\emph{always decreases} with increasing \(R\).

When the substrate is sapphire, the FTM effects for the thermoelastic noises 
are comparable to the Brownian-substrate errors for fused silica. When 
\(R=16\rm{ cm}\), the FTM effects on the sapphire thermoelastic noises 
correspond to a fractional error \(|1-\Delta|\) of 20\% -- 30\%. 

\subsubsection{Switching from a Gaussian beam to a mesa beam with the same 
diffraction loss and mirror radius}
\begin{figure}
\centerline{\includegraphics[width=8cm]{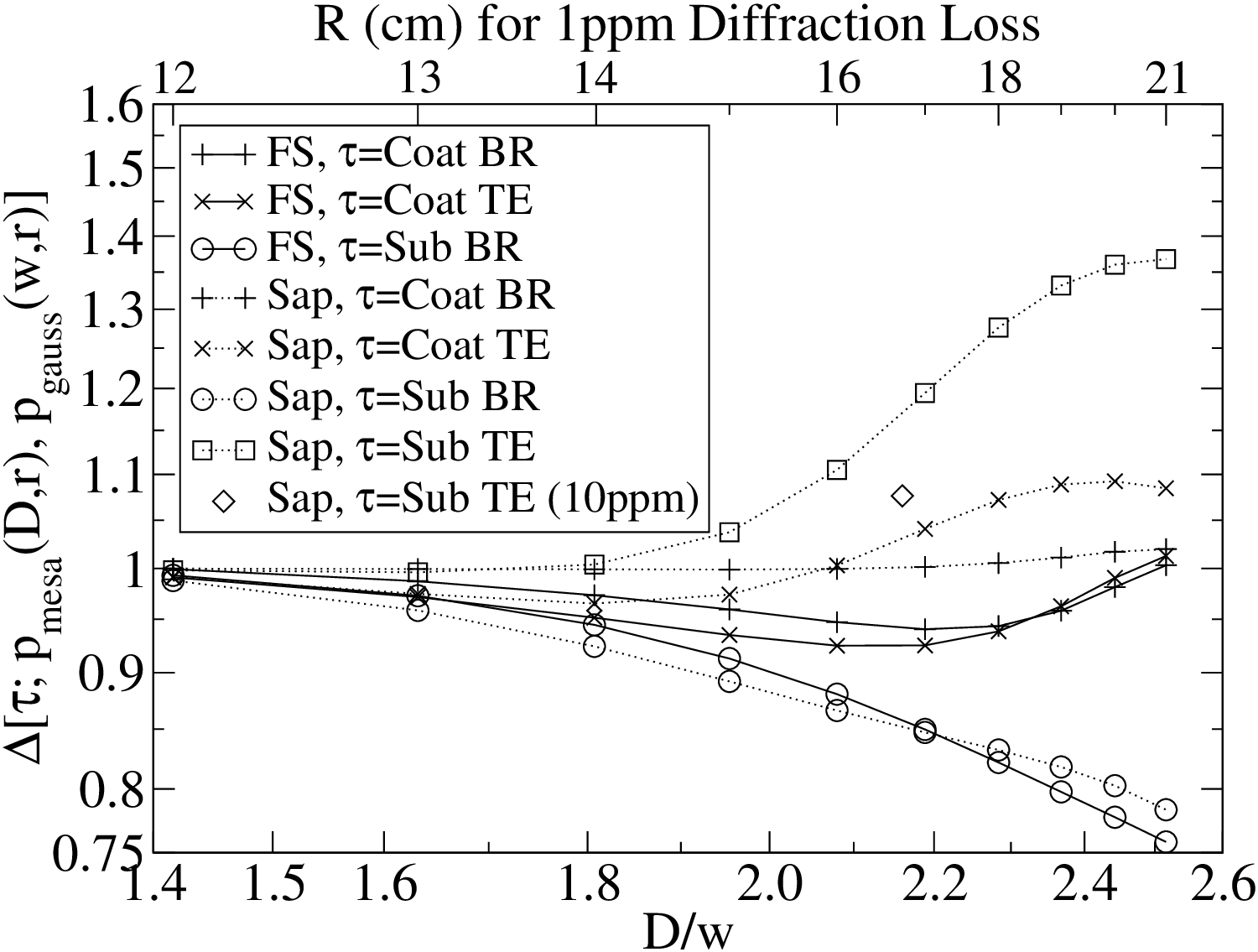}}
\caption{A log-log plot of 
\(\Delta[\tau;p_{\rm{mesa}}(D;r),p_{\rm{gauss}}(w;r)]\). The beam width 
parameters \(w\) and \(D\) are chosen so that the diffraction loss is 1 ppm 
(unless 10 ppm is indicated). The corresponding mirror width for 1 ppm 
diffraction losses is shown on the top axis; the 10 ppm point corresponds to a 
mirror radius of 15.7 cm. The fractional error of the sensitivity change made 
by neglecting edge effects is \(|1-\Delta|\). The FTM values are obtained by 
taking ratios of the noises calculated by Agresti and DeSalvo 
\cite{Agresti:2005b}, except for the 10 ppm value, which is due to 
O'Shaughnessy, Strigin, and Vyatchanin \cite{OSV:2003}. FS and Sap mean 
fused-silica and sapphire substrates. (The fused-silica substrate 
thermoelastic noise is negligible; this case is omitted from the figure.)}
\label{figCFTMGaussToMesa}
\end{figure}
The errors due to neglecting FTM effects in the Gaussian-to-mesa case behave 
qualitatively differently from (and are generally smaller than) the 
resized-beam errors. Figure \ref{figCFTMGaussToMesa} plots 
\(\Delta[\tau;p_{\rm{mesa}}(D;r),p_{\rm{gauss}}(w;r)]\) for fused silica and 
sapphire substrates. For both fused-silica and sapphire substrates, the 
coating sensitivity changes are not strongly sensitive to edge effects; in 
these cases, \(C_{\rm{FTM}}\) and \(C_{\rm{ITM}}\) differ by less than about 
10\% even when the beam widths exceed 17 cm (and thus are significant fractions 
of \(R\) and \(H\) [c.f. figures \ref{figDimlessGauss} and 
\ref{figDimlessMesa}]). The substrate sensitivity changes are more sensitive to 
edge effects, but even then the edge effects remain below about 15\%, provided 
that \(R\lesssim 17\rm{ cm}\) for fused-silica substrates and 
\(R\lesssim 16\rm{ cm}\) for sapphire substrates.
\section{Conclusion}
\label{sec:conclusion} 
Changing the shape of the laser beam in advanced LIGO can reduce the thermal 
noise, which is the limiting noise source at frequencies from 40 Hz to 200 Hz. 
In the Fourier domain, the relations between the thermal noise and the beam 
shape for semi-infinite mirrors take the form of simple scaling laws. Moreover, 
the coating thermal noises obey the same local scaling law. These results 
enable a straightforward comparison of the thermal noises for two different 
beam shapes when edge effects are neglected. The scaling laws predict the 
improvement of mesa-beam sensitivities vs. Gaussian-beam sensitivities quite 
well.  For 40 kg, fused-silica mirrors, the substrate-noise scaling laws agree 
with the finite-mirror results within approximately 15\% for mirror sizes not 
larger than the advanced-LIGO baseline size of about 17 cm; the coating-noise 
scaling laws agree with the finite-mirror predictions to better than about 
10\%. Therefore, the infinite-test-mass scaling laws may be a very useful tool 
for estimating optimal beam shapes for advanced LIGO and other future 
gravitational-wave interferometers.

\ack{
I would like to thank Kip Thorne for suggesting this problem and for his 
advice and encouragement. I would also like to thank Juri Agresti for helpful 
discussions as well as for providing for comparison the data that will be 
published in \cite{Agresti:2005b}. This work was supported in part by 
NSF grants PHY-0099568 and PHY-0601459. The numerical computations described 
in this paper were performed using \emph{Mathematica} version 5.2. The 
figures, including best fit lines, were prepared using Grace 5.1.18.
}

\appendix

\section{Derivation of equations (\ref{eq:apResSub}) and (\ref{eq:apRes})}
\label{sec:nonlocalCancels}
In this appendix, I derive equation (\ref{eq:apResSub}), which I use in the 
derivation of the scaling law (\ref{eq:BRSubRes}) for Brownian substrate noise. 
Then, I deduce equation (\ref{eq:apRes}), which I use 
in the derivation of the scaling law (\ref{eq:BRCoatRes}) for Brownian coating noise.

First, consider the integral
\begin{equation}\label{eq:ap:eq:theIntegral}
\int_0^\infty {\rmd}r r \left[ S_{\varphi\varphi}^2 
- S_{\varphi\varphi} \left(\theta-S_{zz}\right)\right].
\end{equation} Combining equations (\ref{eq:thetaSub}) and (\ref{eq:SzzSub}) gives
\begin{eqnarray}\label{ap:eq:thetaMinusSzz}
\theta - S_{zz} = \frac{1}{2\mu} \int_0^\infty {\rmd}k \rme^{-k z} 
\tilde{p}(k)\left[\frac{-\mu}{\lambda+\mu}+kz\right] k J_0(k r).\nonumber\\ 
\end{eqnarray} Inserting equations (\ref{ap:eq:thetaMinusSzz}) and 
(\ref{eq:SphiphiSub}) into the left hand side of 
equation (\ref{eq:ap:eq:theIntegral}) yields
\begin{eqnarray}\label{ap:eq:expandedEq}
\fl \int_0^\infty {\rmd}r r \left[S_{\varphi\varphi}^2 
- S_{\varphi\varphi} \left(\theta-S_{zz}\right)\right]\nonumber\\ \fl = 
\frac{1}{4\mu^2} \int_0^\infty {\rmd}k \int_0^\infty {\rmd}k^\prime  
\rme^{-(k+k^\prime) z}\left[\frac{-\mu}{\lambda+\mu}+k z\right]
\left[\frac{-\mu}{\lambda+\mu}+k^\prime z\right] \tilde{p}(k) 
\tilde{p}(k^\prime) I,
\end{eqnarray}
where
\begin{equation}\label{ap:eq:I}
I = \int_0^\infty {\rmd}r \frac{J_1(k r) J_1(k^\prime r)}{r} 
- k^\prime \int_0^\infty {\rmd}r J_1(k r) J_0(k^\prime r).
\end{equation} Since \(k\) and \(k^\prime\) are variables of integration, 
and since aside from \(I\) itself, (\ref{ap:eq:expandedEq}) is 
unchanged by letting \(k\leftrightarrow k^\prime\), \(I\) can be rewritten as
\begin{eqnarray}\label{ap:eq:Irewrite}
\fl I =  \int_0^\infty {\rmd}r \frac{J_1(k r) J_1(k^\prime r)}{r} 
- \frac{1}{2} k^\prime \int_0^\infty {\rmd}r J_1(k r) J_0(k^\prime r)
- \frac{1}{2} k \int_0^\infty {\rmd}r J_1(k^\prime r) J_0(k r).
\end{eqnarray} The integrals in (\ref{ap:eq:Irewrite}) are special cases 
of equations (11.4.33), (11.4.34), and (11.4.42) of 
\cite{Abramowitz}:
\numparts
\begin{eqnarray}\label{ap:eq:j1j1int}
\int_0^\infty {\rmd}r \frac{J_1(k r) J_1(k^\prime r)}{r} & = & 
\frac{k^\prime}{2 k} \eta(k-k^\prime) + \frac{k}{2 k^\prime} 
\eta(k^\prime - k),\nonumber\\ \\
\label{ap:eq:j1j0int}
\int_0^\infty {\rmd}r J_1(k r) J_0(k^\prime r) & = & \frac{\eta(k-k^\prime)}{k}.
\end{eqnarray}
\endnumparts Here \(\eta\) is the unit step function. Inserting 
(\ref{ap:eq:j1j1int}) and (\ref{ap:eq:j1j0int}) into 
(\ref{ap:eq:Irewrite}) shows that
\begin{equation}\label{ap:eq:resultSub}
I = 0 \Rightarrow \int_0^\infty {\rmd}r r \left[ S_{\varphi\varphi}^2 
- S_{\varphi\varphi} \left(\theta-S_{zz}\right)\right] = 0, 
\end{equation} which is equation (\ref{eq:apResSub}).

Next, combining equations (\ref{eq:thetaCoat}) -- (\ref{eq:SzzCoat}) shows that
\numparts
\begin{eqnarray}
S_{\varphi\varphi}^{\rm{coat}} = S_{\varphi\varphi}\left.\right|_{z=0}\\
\theta^{\rm{coat}} - S_{zz}^{\rm{coat}} = \left.\left(\theta-S_{zz}\right)
\right|_{z=0}.
\end{eqnarray} 
\endnumparts Thus, setting \(z=0\) in (\ref{ap:eq:resultSub}) gives
\begin{equation}
\int_0^\infty {\rmd}r r \left[ \left(S_{\varphi\varphi}^{\rm{coat}}\right)^2 
- S_{\varphi\varphi}^{\rm{coat}} 
\left(\theta^{\rm{coat}}-S_{zz}^{\rm{coat}}\right)\right] = 0,
\end{equation} which is equation (\ref{eq:apRes}).

\section{Junction conditions for the stress and strain of a statically 
deformed, semi-infinite mirror with thin coating}\label{app:junction}
The junction conditions (\ref{eq:thetaCoat}) -- (\ref{eq:SzzCoat}) are 
listed in equation (A4) of \cite{harry:2002} along with the statement that 
for these conditions to hold, the Poisson ratios of the coating and substrate 
should not be ``too different.'' This restriction is actually unnecessary, 
provided that the coating is sufficiently thin. One can see this as follows:

Because the coating adheres to the substrate surface, the substrate surface 
and coating have the same tangential displacement. Continuity of \(u_r\) and 
\(u_\varphi\) immediately implies continuity of \(S_{rr}\) and 
\(S_{\varphi\varphi}\). A straightforward pillbox integration of the 
equilibrium condition \(\nabla_j T_{ij} = 0\) then shows that \(T_{zz}\) and 
\(T_{rz}\) are also continuous across the junction.

All of the other junction conditions given in equation (A.4) of
\cite{harry:2002} then follow, with one exception: the junction 
condition on \(S_{(rz)}\) should read 
\(\mu^{\rm{coat}} S^{\rm{coat}}_{(rz)} = \mu^{\rm{sub}} S^{\rm{sub}}_{(rz)}\), 
not \(S^{\rm{coat}}_{(rz)} = S^{\rm{sub}}_{(rz)}\). But since \(T_{rz}=0\) on 
the coating surface (and thus also to high accuracy throughout the thin 
coating), this error is irrelevant; the correct junction condition is simply 
\(S^{\rm{coat}}_{(rz)} = S^{\rm{sub}}_{(rz)}=0\).

\section*{References}
\bibliographystyle{iopart-num}
\bibliography{lovelaceLIGOThermalNoise}

\providecommand{\newblock}{}
\begin{thebibliography}{10}
\expandafter\ifx\csname url\endcsname\relax
  \def\url#1{{\tt #1}}\fi
\expandafter\ifx\csname urlprefix\endcsname\relax\def\urlprefix{URL }\fi
\providecommand{\eprint}[2][]{\url{#2}}

\bibitem{AdvancedLIGOProposal}
Advanced LIGO: context and summary (online document accessed 14 September,
  2006)
  \urlprefix\url{http://www.ligo.caltech.edu/advLIGO/scripts/summary.shtml}

\bibitem{Agresti:2004}
Agresti J 2005 Researches on non-standard optics for advanced G.W.
  interferometers (internal LIGO document LIGO-T040225-00-R)
  \urlprefix\url{http://www.ligo.caltech.edu/docs/T/T040225-00.pdf}

\bibitem{liuThorne:2000}
Liu Y~T and Thorne K~S 2000 {\em Phys. Rev. D\/} {\bf 62} 122002

\bibitem{harry:2002}
Harry G~M {\em et~al.\/} 2002 {\em Class. Quantum Grav.\/} {\bf 19} 897

\bibitem{BV:2003}
Braginsky V~B and Vyatchanin S~P 2003 {\em Phys. Lett. A\/} {\bf 312} 244

\bibitem{AOT}
d'Ambrosio E, O'Shaughnessy R and Thorne K~S 2000 Beam reshaping to reduce
  thermal noise (internal LIGO document G000223-00-D)
  \urlprefix\url{http://www.ligo.caltech.edu/docs/G/G000223-00.pdf}

\bibitem{AOSTV}
D'Ambrosio E {\em et~al.\/} 2004  (submitted to Phys. Rev. D.)
  (\textit{Preprint} \eprint{gr-qc/0409075})

\bibitem{OSV:2003}
O'Shaughnessy R, Strigin S and Vyatchanin S 2003  (submitted to Phys. Rev. D)
  (\textit{Preprint} \eprint{gr-qc/0409050})

\bibitem{Agresti:2005a}
Agresti J and DeSalvo R 2005 flat beam profile to depress thermal noise
  (internal LIGO document G050041-00-Z)
  \urlprefix\url{http://www.ligo.caltech.edu/docs/G/G050041-00}

\bibitem{Agresti:2005b}
Agresti J and DeSalvo R 2007 Document in preparation.

\bibitem{Agrestietal:2006}
Agresti J {\em et~al.\/} 2006 {\em J. Phys: Conference Series\/} {\bf 32} 301

\bibitem{Taralloetal:2007}
Tarallo M {\em et~al.\/} 2007 Generation of a flat-top laser beam for
  gravitational-wave detectors by means of a non-spherical fabry-perot
  resonator (submitted to Applied Optics)
  \urlprefix\url{http://www.ligo.caltech.edu/$\sim$jmiller/Documents/MHpaper.p%
df}

\bibitem{levin:1998}
Levin Y 1997 {\em Phys Rev. D\/} {\bf 57} 659

\bibitem{fejeretal:2004}
Fejer M~M {\em et~al.\/} 2004 {\em Phys. Rev. D\/} {\bf 70} 082003

\bibitem{OShaughnessy:2006v1}
O'Shaughnessy R 2006  (\textit{Preprint} \eprint{gr-qc/0607035v1})

\bibitem{OShaughnessy:2006}
O'Shaughnessy R 2006 {\em Class. Quantum Grav.\/} {\bf 23} 7627

\bibitem{Vyatchanin:2004}
Vyatchanin S 2004 Fdt approach calculations of brownian noise in thin layer
  (internal ligo document t040242-00z)
  \urlprefix\url{http://www.ligo.caltech.edu/docs/T/T040242-00.pdf}

\bibitem{vinet:2005}
Vinet J~Y 2005 {\em Class. Quantum Grav.\/} {\bf 22} 1395

\bibitem{BHV}
Bondu F, Hello P and Vinet J~Y 1998 {\em Phys. Lett. A\/} {\bf 246} 227

\bibitem{Siegman:1977}
Siegman A~E 1977 {\em Opt. Lett.\/} {\bf 1} 13

\bibitem{AdvancedLIGOProposalOld}
Fritschel P 2001 Advanced ligo systems design (internal ligo document
  t010075-00-d)
  \urlprefix\url{http://www.ligo.caltech.edu/docs/T/T010075-00.pdf}

\bibitem{Abramowitz}
Abramowitz M and Stegun I~A (eds) 1964 {\em Handbook of Mathematical
  Functions\/} (New York: Dover Publications)

\end{thebibliography}

\end{document}